\documentclass[twocolumn]{aastex63}
\usepackage{graphicx}
\usepackage{amsmath}
\usepackage{multirow}

\newcommand{\as}{^{\prime \prime}}
\newcommand{\kms}{\rm km~s^{-1}}
\newcommand{\ceo}{\rm C^{\rm 18}O}
\newcommand{\mlin}{M_{\rm line}}
\newcommand{\mspc}{M_{\odot} ~\rm pc^{-1}}
\received{February 22, 2023}
\revised{April 26, 2023}
\accepted{May 11, 2023}
\shorttitle{B-fields of the Filaments in the Cal-X Hub}
\shortauthors{Chung et al.}

\begin{document}

\title{Magnetic Fields and Fragmentation of Filaments in the Hub of California-X}

\author{Eun Jung Chung}
\affiliation{Department of Astronomy and Space Science, Chungnam National University, Daejeon, Republic of Korea}

\author{Chang Won Lee} \affiliation{Korea Astronomy and Space Science Institute, 776 Daedeokdae-ro, Yuseong-gu, Daejeon 34055, Republic of Korea} \affiliation{University of Science and Technology, Korea (UST), 217 Gajeong-ro, Yuseong-gu, Daejeon 34113, Republic of Korea}

\author{Woojin Kwon} \affiliation{Department of Earth Science Education, Seoul National University, 1 Gwanak-ro, Gwanak-gu, Seoul 08826, Republic of Korea} \affiliation{SNU Astronomy Research Center, Seoul National University, 1 Gwanak-ro, Gwanak-gu, Seoul 08826, Republic of Korea}

\author{Mario Tafalla} \affiliation{Observatorio Astron\'omico Nacional (IGN), Alfonso XII 3, E-28014 Madrid, Spain}

\author{Shinyoung Kim} \affiliation{Korea Astronomy and Space Science Institute, 776 Daedeokdae-ro, Yuseong-gu, Daejeon 34055, Republic of Korea}

\author{Archana Soam} \affiliation{Indian Institute of Astrophysics, II Block, Koramangala, Bengaluru 560034, India}

\author{Jungyeon Cho} \affiliation{Department of Astronomy and Space Science, Chungnam National University, Daejeon, Republic of Korea}

\begin{abstract}
We present 850~$\mu$m polarization and $\ceo~(3-2)$ molecular line observations toward the X-shaped nebula in the California molecular cloud using the JCMT SCUBA-2/POL-2 and HARP instruments. The 850~$\mu$m emission shows that the observed region includes two elongated filamentary structures (Fil1 and Fil2) having chains of regularly spaced cores. We measured the mass per unit length of the filament and found that Fil1 and Fil2 are thermally super- and subcritical, respectively, %both filaments are thermally super/transcritical
but both are subcritical if nonthermal turbulence is considered. The mean projected spacings~($\Delta\bar S$) of cores in Fil1 and Fil2 are 0.13 and 0.16~pc, respectively. $\Delta\bar S$ are smaller than $4\times$filament width expected in the classical cylinder fragmentation model. The large-scale magnetic field orientations shown by Planck are perpendicular to the long axes of Fil1 and Fil2, while those in the filaments obtained from the high-resolution polarization data of JCMT are disturbed, but those in Fil1 tend to have longitudinal orientations. Using the modified Davis-Chandrasekhar-Fermi (DCF) method, we estimated the magnetic field strengths ($B_{\rm pos}$) of filaments which are 110$\pm$80 and 90$\pm$60~$\mu$G. We calculated the gravitational, kinematic, and magnetic energies of the filaments, and found that the fraction of magnetic energy is larger than 60\% in both filaments. We propose that a dominant magnetic energy may lead the filament to be fragmented into aligned cores as suggested by Tang et al., and a shorter core spacing can be due to a projection effect via the inclined geometry of filaments or due to a non-negligible, longitudinal magnetic fields in case of Fil1.
\end{abstract}


\section{Introduction}

Hub-filament systems (HFSs) are the best laboratories to investigate the initial conditions for star formation. HFS consists of a hub with high column density ($> 10^{22}~\rm cm^{-2}$) and low axis ratio and several filaments with relatively low column density and high aspect ratio extended from the hub \citep{myers2009}. They are mostly associated with active low- to high-mass star clusters \citep{kumar2020}, and easily found both in nearby star-forming molecular clouds and more distant infrared dark clouds. Hence, HFSs have been studied using multi-wavelength observations to understand how they form and stars are generated in them \citep[e.g.,][]{kumar2020,hwang2022,bhadari2022}.

The crucial process to form stars in the hubs and filaments is the fragmentation. It is believed that early star formation begins with fragmentation of filaments in hydrostatic equilibrium state into cores due to linear perturbations \citep[e.g.,][]{ostriker1964}. With an assumption of the isothermal, infinitely long cylindrical structure, fragmentation in a filament occurs via gravitational perturbations with a critical wavelengths of 2 times the filament's diameter, and filaments may have cores with a regular spacing of 4 times the diameter of filament which is the fastest growing mode \citep[e.g.,][]{inutsuka1992}. However, the observed cores' separations are generally not matched to the spacing expected by the classical cylinder model \citep[e.g.,][]{tafalla2015,zhang2020}, probably because of the other factors which can affect the fragmentation process of filaments such as turbulence, accreting flows, and/or magnetic fields \citep[e.g.,][]{fiege2000i,clarke2016,hanawa2017}. 

%=========== FIGURE : Obs. region. Herschel 250 micron image of L1478 F8/9/10
\begin{figure*} \epsscale{1.17}
\includegraphics[width=1\textwidth]{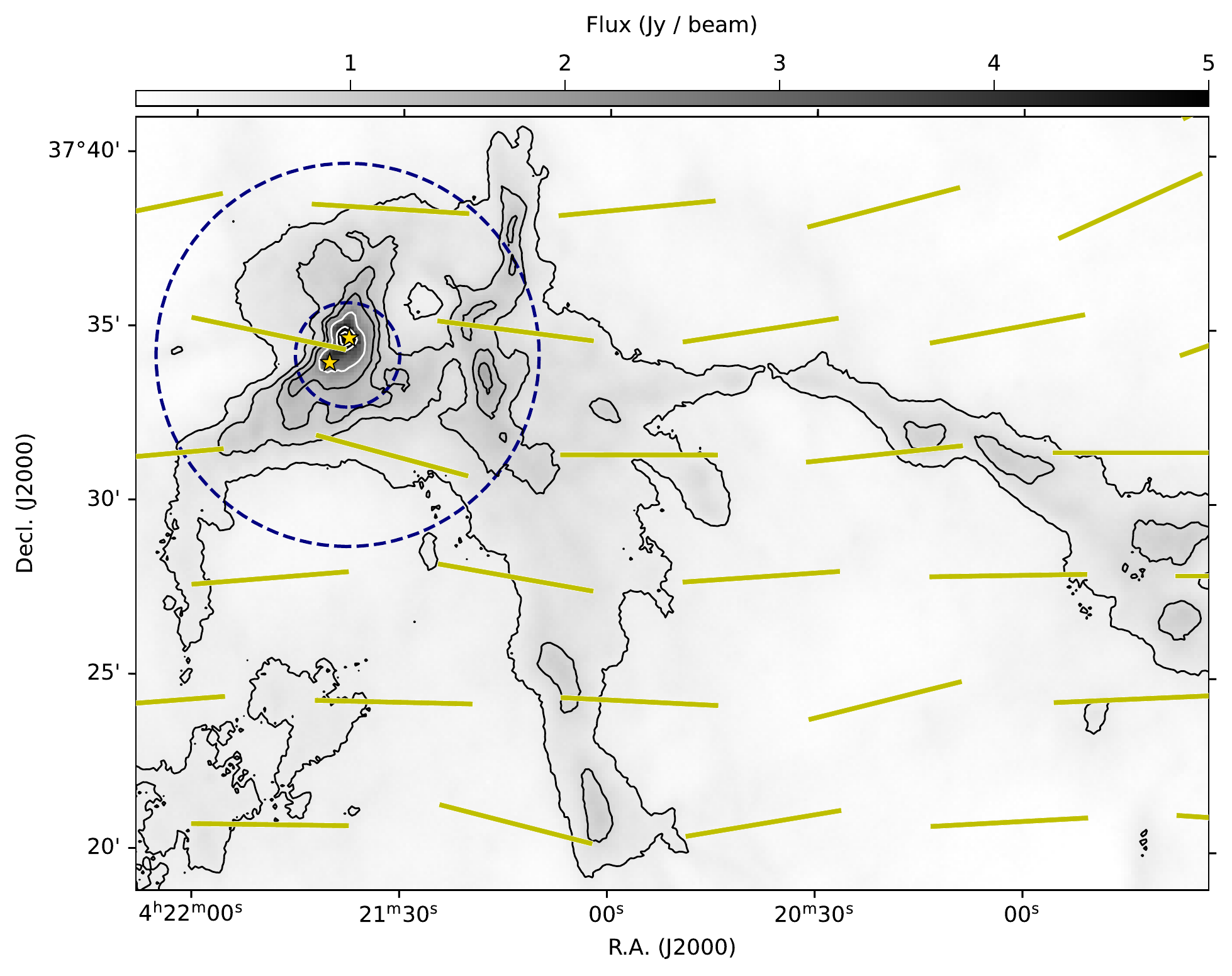}
\caption{Herschel 250~$\mu$m image of the X-shaped nebula region in the California molecular cloud. The contour levels are 3, 6, 9, 12, 20, 40, and 70$\times \sigma$ (1$\sigma$ = 0.12~Jy~beam$^{-1}$). The yellow segments depict the large scale magnetic field orientations obtained by rotating the submillimeter Planck 353~GHz polarization orientations by 90~degree. The effective angular resolution is $\sim 5^{\prime}$. The yellow stars denote YSOs found by \citet{harvey2013}. The JCMT SCUBA-2/POL-2 observing area of 11$^{\prime}$ diameter is indicated with the large dashed circle. The small dashed circle shows the inner 3$^{\prime}$ region with the best sensitivity. \label{fig:obsregion}}
\end{figure*}
%=========== 

The main drivers of star formation are gravity, turbulence, and magnetic fields, although their precise roles, particularly during the fragmentation process from filamentary molecular clouds into dense cores, are still unclear. Recently, it has been proposed that the relative significance of these three factors can determine the different evolutionary paths from clumps on the scale of 2 pc to cores on the scale of 0.6 pc \citep{tang2019}. Specifically, \citet{tang2019} has classified fragmentation types into $clustered$, $aligned$, and $no$ fragmentation based on the distribution of cores within the natal clouds, with each type appearing to be closely related to the dominance of turbulence, magnetic fields, and gravity, respectively. However, more observational data on various filamentary molecular clouds with different fragmentation types are needed to better understand the precise roles of gravity, turbulence, and magnetic fields in star formation.

L1478 in the California molecular cloud is known as a low-mass star-forming cloud at a distance of 470~pc \citep{zucker2019}. It has a prominent HFS to which \citet{imara2017} refer to as California-X (shortly Cal-X) because of its X-shape. The Herschel 250~$\mu$m image of Cal-X given in Figure~\ref{fig:obsregion} shows that there are two long pc-scale filaments radiating from the bright hub to the south and to the west, respectively. The mass of the hub is $\sim 130 ~M_{\odot}$ \citep{chung2019}, and those of filaments at the south and west are $\sim 130$ and $150~M_{\odot}$, respectively \citep{imara2017}. The hub includes two YSOs: one is Class I and the other is Class II \citep{harvey2013,broekhoven-fiene2014}. The continuous velocity gradients of Cal-X indicate possible gas flow along the filaments into the hub \citep{imara2017,chung2019}. The Planck data show, though its resolution is limited ($\sim 5^{\prime}$), that magnetic field orientations are mostly east-to-west, hence the long filament in the west is roughly parallel to the global B-field but the hub and the southern filament is perpendicular to the global B-field. 

At the central $11^{\prime}$ area of the hub, two elongated filamentary features can be seen. \citet{zhang2020} investigated these filaments and dense cores in the hub of Cal-X. They showed that the cores are regularly spaced along the filaments where the core spacings are shorter than the expected spacing by the classical cylinder model \citep{inutsuka1992}. We notice that the chain of cores in the filaments of the Cal-X hub is classified as $aligned$ fragmentation, and thus the filaments are suitable to study the role of gravity, turbulence, and magnetic fields on the fragmentation of hub/filament into cores. We have performed high-resolution polarization observations and molecular line observations using the SCUBA-2/POL-2 and HARP instruments mounted on the JCMT toward the hub of Cal-X. The paper is organized as follows. In Section~\ref{sec:obsdr}, we describe the observations and data reductions. The results of the observations and the measured magnetic field strength are depicted in Section~\ref{sec:results}. We present the analysis and discussion in Sections~\ref{sec:anal} and \ref{sec:disc}, respectively. A summary is given in Section~\ref{sec:summ}. \\

\section{Observations} \label{sec:obsdr}

\subsection{Polarization Observations}

We made submillimeter continuum and polarization observations at 850~$\mu$m toward the hub of the California-X molecular cloud. The observation was performed with the SCUBA-2/POL-2 instrument on the James Clerk Maxwell Telescope (JCMT) between 2019 October and 2021 January. The beam size at 850~$\mu$m wavelength is 14$\as$.1 (corresponding to $\sim$0.03~pc at a distance of 470~pc). The standard SCUBA-2/POL-2 daisy mapping mode was used with a constant scanning speed of 8$\as~ \rm s^{-1}$. The observations were done with 21~times, an average integration time of 40~minutes under dry weather conditions with submillimeter opacity at 225~GHz ($\tau_{\rm 225~GHz}$) ranging between 0.05 and 0.08. 

%=========== FIGURE : 850 micron image and filaments identification
\begin{figure*} \epsscale{1.17}
\plotone{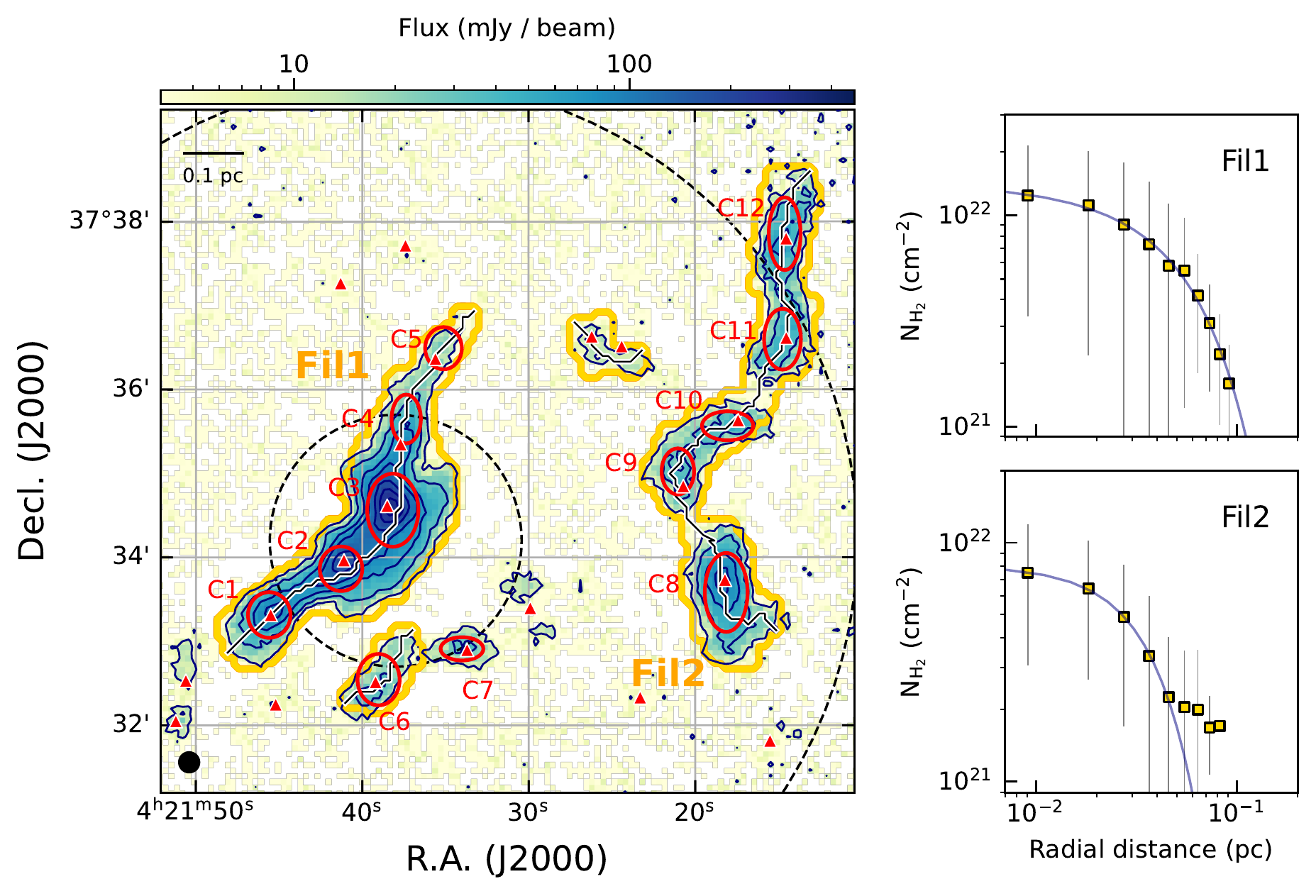}
\caption{The observed 850~$\mu$m Stokes {\em I} image and contours. The contour levels are 3, 10, 20, 30, 50, 70, and 90$\times \sigma$ (1$\sigma$ is 3.2~mJy~beam$^{-1}$). Filaments identified with {\sc filfinder} are presented with yellow polygons. The filaments' skeletons are drawn with solid lines. The red ellipses depict the 850~$\mu$m cores identified using  {\sc FellWalker} and the red triangles indicate the positions of cores identified with the Herschel data \citep{zhang2020}. The dashed circles are the observing area of 11$^{\prime}$ diameter and the best sensitivity coverage of 3$^{\prime}$ region. The black circle at the bottom left corner shows the POL-2 850~$\mu$m beam size of 14.1$\as$. A reference scale of 0.1~pc is shown on the top left corner. {\it Right:} Averaged radial column density profiles of filament1 and 2 centered on their skeletons (yellow squares) and their Gaussian fits to estimate the filaments' widths. \label{fig:filfind}}
\end{figure*}
%=========== 

We used the {\tt pol2map} script of the STARLINK/SMURF package for the 850~$\mu$m data reduction. The {\tt pol2map} data reduction process consists of three steps. In the first step, the raw bolometer time streams for each observation are converted into separate Stokes {\em I}, {\em Q}, and {\em U} time streams using the process {\it calcqu}. In the second step, it produces improved {\em I} maps  using a mask determined with signal-to-noise ratio via the process {\it makemap}. We set the parameter {\tt SKYLOOP=TRUE} to reduce the dispersion between maps and lessen the intrinsic instabilities of the map-making algorithm\footnote{\url{http://starlink.eao.hawaii.edu/docs/sc22.htx/sc22.html}}. The final {\em I} map is created by co-adding the improved individual {\em I} maps. In the final step, {\em Q} and {\em U} maps are produced from the {\em Q} and {\em U} time streams with the same masks used in the previous step. For the instrumental polarization correction, the `August 2019' IP model\footnote{\url{https://www.eaobservatory.org/jcmt/2019/08/new-ip-models-for-pol2-data/}} was used. The final {\em I}, {\em Q}, and {\em U} maps are binned with a pixel size of 4$\as$. 

The polarized intensity ({\em PI}) is the quadratic sum of {\em Q} and {\em U}, $PI=\sqrt{Q^{2} + U^{2}}$, and thus the noises of {\em Q} and {\em U} always make a positive contribution to the polarization intensity \citep[e.g.,][]{vaillancourt2006}. The debiased polarization intensity is estimated using the modified asymptotic estimator \citep{plaszczynski2014}:
\begin{equation}
	PI = \sqrt{Q^{2} + U^{2}} - \sigma^{2} \frac{1 - e^{-(Q^{2} + U^{2})/\sigma^{2}}}{2\sqrt{Q^{2} + U^{2}}},
\end{equation}
where $\sigma^{2}$ is the weighted mean of the variances on {\em Q} and {\em U}:
\begin{equation}
	\sigma^{2} = \frac{Q^{2} \sigma_{Q}^{2} + U^{2} \sigma_{U}^{2}}{Q^{2} + U^{2}},
\end{equation}
and $\sigma_{Q}$ and $\sigma_{U}$ are the standard errors in {\em Q} and {\em U}, respectively.

The debiased polarization fraction $P$ is calculated as
\begin{equation}
	P = \frac{PI}{I}  \label{eq:p},
\end{equation}
and its corresponding uncertainty is
\begin{equation}
	\sigma_{P} = \sqrt{\frac{\sigma^{2}}{I^{2}} + \frac{\sigma_{I}^{2}(Q^{2} + U^{2})}{I^{4}}},
\end{equation}
where $\sigma_{\rm I}$ is the standard error in {\em I}.

We used the final debiased polarization vector catalog provided with a bin-size of 12$\as$ to increase the signal-to-noise ratios in the polarization data. The 12$\as$ bin size is also close to the beam size of the JCMT/POL-2 at 850~$\mu$m, which is 14.1$\as$. The selection criteria of the polarization measurements are set to be (1) the signal-to-noise ratio (S/N) of total intensity larger than 10 ($I / \sigma_{I} > 10$) and (2) the polarization fraction larger than 2 times of its uncertainty ($P / \sigma_{P} > 2$). 

A Flux Calibration Factor (FCF) of 668~Jy~pW$^{-1}$~beam$^{-1}$ is used for the 850~$\mu$m Stokes {\em I}, {\em Q}, and {\em U} data. This FCF is larger than the standard 850~$\mu$m SCUBA-2 flux conversion factor of 495~Jy~pW$^{-1}$~beam$^{-1}$ because a correction factor of 1.35 is multiplied due to the additional losses from POL-2 \citep{dempsey2013,friberg2016,mairs2021}. The rms noise values in the {\em I}, {\em Q}, and {\em U} binned to pixel size 12$\as$ are 3.2, 3.0, and 3.0~mJy~beam$^{-1}$, respectively. \\

\subsection{$C^{18}O$~(3-2) Observations}

We performed $\ceo~(3-2)$ line observations using the JCMT Heterodyne Array Receiver Programme \citep[HARP;][]{buckle2009} to estimate the velocity dispersion of the region. The data were taken as basket-weaved scan maps over 3 nights between January 2020 and July 2022 in weather band~2 ($\tau_{\rm 225~GHz} \sim 0.065 - 0.08$). The spatial resolutions is about 14 arcsec which is the same as that of JCMT/POL-2 850~$\mu$m data, and the spectral resolution is $\sim 0.05~\kms$. The total observing time is $\sim$6 hours. We reduced the data using the ORAC-DR pipeline in STARLINK software \citep{buckle2012} with a recipe of $\rm 'REDUCE\_SCIENCE\_NARROWLINE'$ and obtained the data cube with a 14$\as$ pixel size. We resampled the data cube to a channel width of 0.1~$\kms$ using a 1-d Gaussian kernel. The mean rms level of the final data cube is about 0.06~K[$T_{A}^{\ast}$]. \\

\section{Results} \label{sec:results}

\subsection{Identification of Filaments and cores} \label{ssec:idf}

The 850~$\mu$m Stokes {\em I} map is presented in Figure~\ref{fig:filfind}. The 850~$\mu$m emission closely matches the Herschel 250~$\mu$m emission of the hub presented in Figure~\ref{fig:obsregion}. There are two elongated filamentary structures: one at the center and the other at the west. 

%=========== FIGURE : 850 micron cores on image with YSOs
\begin{figure} %\epsscale{1.17}
\includegraphics[width=0.49\textwidth]{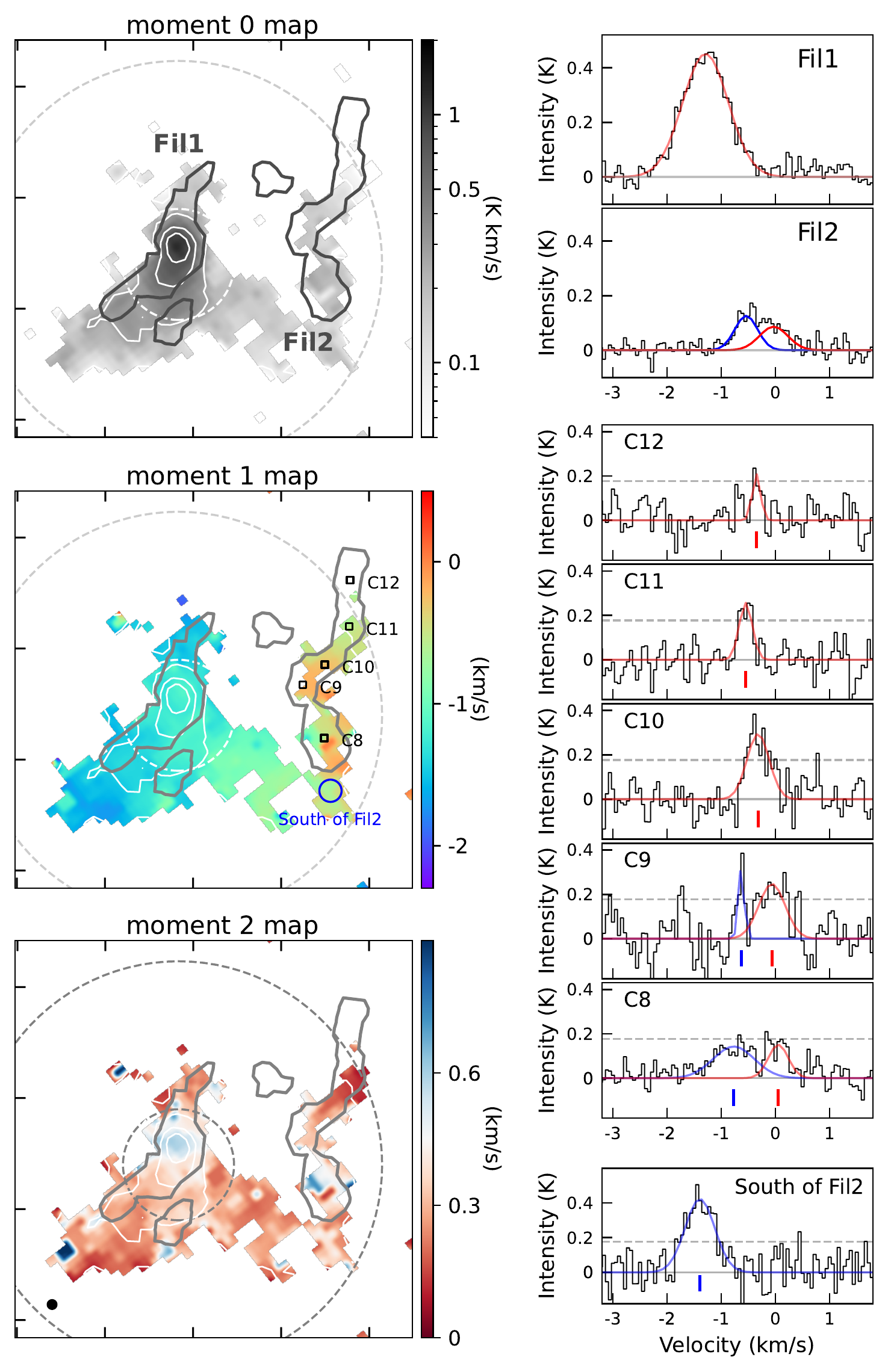}
\caption{{\it Left:} $\ceo~(3-2)$ moment maps. The contours of $\ceo~(3-2)$ integrated intensity are overlaid on the moment maps depicted with gray or color tones, and the contour levels are 3, 10, 20, and 30$\times \sigma$ (1$\sigma$ = 0.03~K~$\kms$). The outlines of filaments are drawn with solid polygons. The dashed circles present the POL-2 observation area of 11$^{\prime}$ diameter and its best sensitivity coverage of 3$^{\prime}$ region. The open squares of moment~1 map depict the position of dense cores in Fil2. The black circle at the bottom left of moment~2 map shows the FWHM beam size at the $\ceo~(3-2)$ frequency. {\it Right:} the averaged spectra of filaments and dense cores. Red profiles overlaid on the spectra are the Gaussian fit results of filaments and dense cores, and the blue profiles are the second Gaussian components. The spectrum shown in the bottom panel is the averaged one of the southern region of Fil2 depicted with blue circle on the moment~1 map, i.e., printed as `South of Fil2'. \label{fig:c18o}}
\end{figure}
%===========

We used the {\sc filfinder} algorithm, which employs mathematical morphology to identify filaments \citep{koch2015}. {\sc filfinder}  takes five steps to identify filamentary structures. Simply introducing the algorithm here, it first flattens the image using an arctan transform of $I^{\prime} = I_{0} \rm arctan(\it I/I_{0})$ with the normalization of $I_{0} \equiv \rm exp(\mu + 2 \sigma)$ where $\mu$ and $\sigma$ are the mean and standard deviation of the log-intensity. Secondly, it makes the flattened data to be smoothed with a Gaussian beam having a full width half maximum (FWHM) of 0.05~pc. And then, it creates a mask using an adaptive threshold of the smoothed data, i.e., it keeps a pixel which has a greater intensity than the median value of the neighboring pixels within the distance of 0.1~pc, while discarding a pixel having a lower intensity. In the forth and fifth steps, small and spurious structures are removed, i.e., structures with sizes less than $5 \pi \rm (0.1~pc)^{2}$ are rejected, and small spurious features of the edges are also removed by applying a  0.05~pc size median filter. 

% TABLE : filament identified
\begin{deluxetable*}{lccccccccccc} \tablecaption{Derived Physical Parameters of the Filaments \label{tab:ppfila}}
%\tablecolumns{14}
%\tablewidth{0pt}
\tablehead{
\colhead{} &
\colhead{Length} &
\colhead{Width} &
\colhead{$N_{\rm H_{2}}^{0}$} &
\colhead{$\bar n_{\rm H_{2}}$} &
\colhead{$M$} & 
\colhead{$M_{\rm line}$} &
\colhead{$\sigma_{\rm NT}$} \vspace{-2mm}
\tabularnewline  
\colhead{} &
\colhead{(pc)} &
\colhead{(pc)} &
\colhead{($\rm 10^{21}~cm^{-2}$)} &
\colhead{($\rm 10^{3}~cm^{-3}$)} &
\colhead{($M_{\odot}$)} &
\colhead{($M_{\odot}~\rm pc^{-1}$)} &
\colhead{($\kms$)}
}
\startdata
%Fil1 & 0.73 & 0.075$\pm$0.002 & 15$\pm$2 & 66$\pm$10 & 15$\pm$2 & 20$\pm$3 & 0.41\\ 
%F\#10$^{\dagger}$ & 0.52 & 0.08 & 12$\pm$1.9 & 38.5 $\pm$7.7 & 13 & 25 & \\ 
Fil1 & 0.73$\pm$0.04 & 0.160$\pm$0.026 & 13$\pm$8 & 26$\pm$16 & 15$\pm$2 & 20$\pm$3 & 0.41\\ %\hline
%Fil2 & 0.89 & 0.046$\pm$0.002 & 12$\pm$2 & 82$\pm$14 & 8$\pm$1 & 9$\pm$2 & 0.24\\ 
%F\#8$^{\dagger}$ &  0.92 & 0.12 & 9.5$\pm$1 & 22$\pm$3 & 26 & 28 & \\ 
Fil2 & 0.89$\pm$0.05 & 0.070$\pm$0.004 & 7$\pm$5 & 33$\pm$21 & 8$\pm$1 & 9$\pm$2 & 0.24\\ 
\enddata
%\vspace{2mm}
\tablecomments{$M$ is the filament's mass estimated from Equation~\ref{eq:m850}, $N_{\rm H_{2}}^{0}$ is the median column density along the crest of filament, $\bar n_{\rm H_{2}}$ is the average volume density given by $N_{\rm H_{2}}^{0} / W$ by assuming that each filament is cylindrical, and $M_{\rm line}$ is the mass per unit length measured by dividing the mass by its length. } \end{deluxetable*} 
%===========

Using {\sc filfinder}, we obtained four filamentary structures as outlined with yellow in Figure~\ref{fig:filfind}. The filaments' skeletons given by the algorithm are depicted with solid lines. The skeletons are found using a Medial Axis Transform in which the chosen skeleton pixels are the centers of the inscribed circles of the mask. Then, the length of a filament is measured along the longest path through the skeleton after pruning the sub-structures. Among the identified four filaments, we make analyses for the two largest filaments, named filament~1 (Fil1) and filament~2 (Fil2), in this study. 

The mass of filament is estimated with the following equation \citep[e.g.,][]{hildebrand1983}:
\begin{equation}
M = \frac{S_{\nu} ~d^{2}}{\kappa_{\nu}~ B_{\nu}(T_{\rm d})}, \label{eq:m850}
\end{equation}
\noindent where $S_{\nu}$, $\kappa_{\nu}$, and $B_{\nu}$ are the integrated flux density, opacity, and Planck function at the wavelength of 850~$\mu$m, respectively. $T_{\rm d}$ and $d$ are the dust temperature and the distance, respectively. The dust opacity is obtained by $\kappa_{\nu} = 0.1(\nu / 10^{12} \rm Hz)^{\beta} \rm cm^{2}~g^{-1}$ with the assumption of a dust-to-gas ratio of 1:100 \citep{beckwith1991}, and the dust opacity index of $\beta=2$ \citep{draine1984}. The dust temperature was taken from Herschel data \citep{andre2010,chung2019}. The applied $T_{\rm d}$ for Fil1 and Fil2 are 12.0$\pm$1.1~K and 10.8$\pm$0.2~K, and then their masses were derived to be 15$\pm$2 and 8$\pm1~M_{\odot}$, respectively. The H$_{2}$ column density is calculated by dividing the mass in each pixel estimated from the Equation~\ref{eq:m850} by the pixel area. The central H$_2$ column densities ($N_{\rm H_{2}}^{0}$; the median value of $N_{\rm H_{2}}$ along the filament crest) are $13\times 10^{21}$ and 7$\times 10^{21}~\rm cm^{-2}$ for Fil1 and Fil2, respectively. The filaments' widths are estimated from the Gaussian fit of the averaged radial column density profiles as shown in the Figure~\ref{fig:filfind}. The mass per unit length ($\mlin$) is estimated by dividing the mass by the length. $\mlin$ of Fil1 and Fil2 are 20$\pm 3$ and 9$\pm2 ~\mspc$, respectively. The physical properties of Fil1 and Fil2 are listed in Table~\ref{tab:ppfila}. 

\citet{zhang2020} investigated the Cal-X using the Herschel H$_2$ column density map. They used the {\it getfilaments} and {\it getsources} algorithms \citep{menshchikov2012} to identify filaments and dense cores. Filament \#10 and \#8 of \citet{zhang2020} correspond to Fil1 and Fil2 of this study, respectively. Fil1 is longer and wider than F\#10, but Fil2 is shorter and narrower than F\#8. One noticeable thing is that the mass of F\#8 is 26~$M_{\odot}$ at the distance of 470~pc which is about three times larger than Fil2. Besides, the measured line mass of F\#8 is 28~$\mspc$ implying that it is thermally supercritical while Fil2 is subcritical. 

The differences can be caused by the different methods used for identification of filament and measurements of quantities as well as the Herschel far-infrared data from 70~$\mu$m to 500~$\mu$m wavelength and the JCMT sub-mm data at 850~$\mu$m. In addition, we cannot rule out the possibility of underestimation due to the filter-out of the structures with scale greater than a few arcmin and/or decreasing sensitivity at the larger radii than 3$^{\prime}$ of the POL-2 map obtained by the $daisy$ scan mode \citep[e.g.,][]{holland2013}. However, $N_{\rm H_{2}}^{0}$ of Fil1 and Fil2 are consistent with those of F\#10 (12$\times 10^{21}~\rm cm^{-2}$) and F\#8 (9.5$\times 10^{21}~\rm cm^{-2}$). Besides, the average volume densities $\bar n_{\rm H_{2}}$, the key physical quantity used to calculate the magnetic field strength ($B_{\rm POS}$) of Fil1 and Fil2, also agree to those of F\#10 (39$\times 10^{3}~\rm cm^{-3}$) and F\#8 (22$\times 10^{3}~\rm cm^{-3}$) within the uncertainties. 

We used \textsc{FellWalker} clump-finding algorithm \citep{berry2015} to extract dense cores in the filaments. Pixels with intensities $> 1 \sigma$ are used to find cores, and an object having a peak intensity higher than 10$\sigma$ and a size larger than 2$\times$beam size of 14$\as$ is identified as a real core. {\sc FellWalker} algorithm considers the neighboring peaks are separated if the difference between the peak values and the minimum value (dip value) between the peaks is larger than the given threshold. We use 0.9$\sigma$ as the threshold, and found five dense cores each in Fil1 and Fil2 regions, respectively, as shown in Figure~\ref{fig:filfind}. The dense cores identified from the Herschel H$_{2}$ column density map \citep{zhang2020} are presented with red triangles. The positions of 850~$\mu$m cores are consistent with those of Herschel dense cores. C4, C5, and C9 have offsets, but within one beam size of JCMT. 

%=========== FIGURE : Polarization vectors on 850 micron image
\begin{figure*} \epsscale{1.17}
\plotone{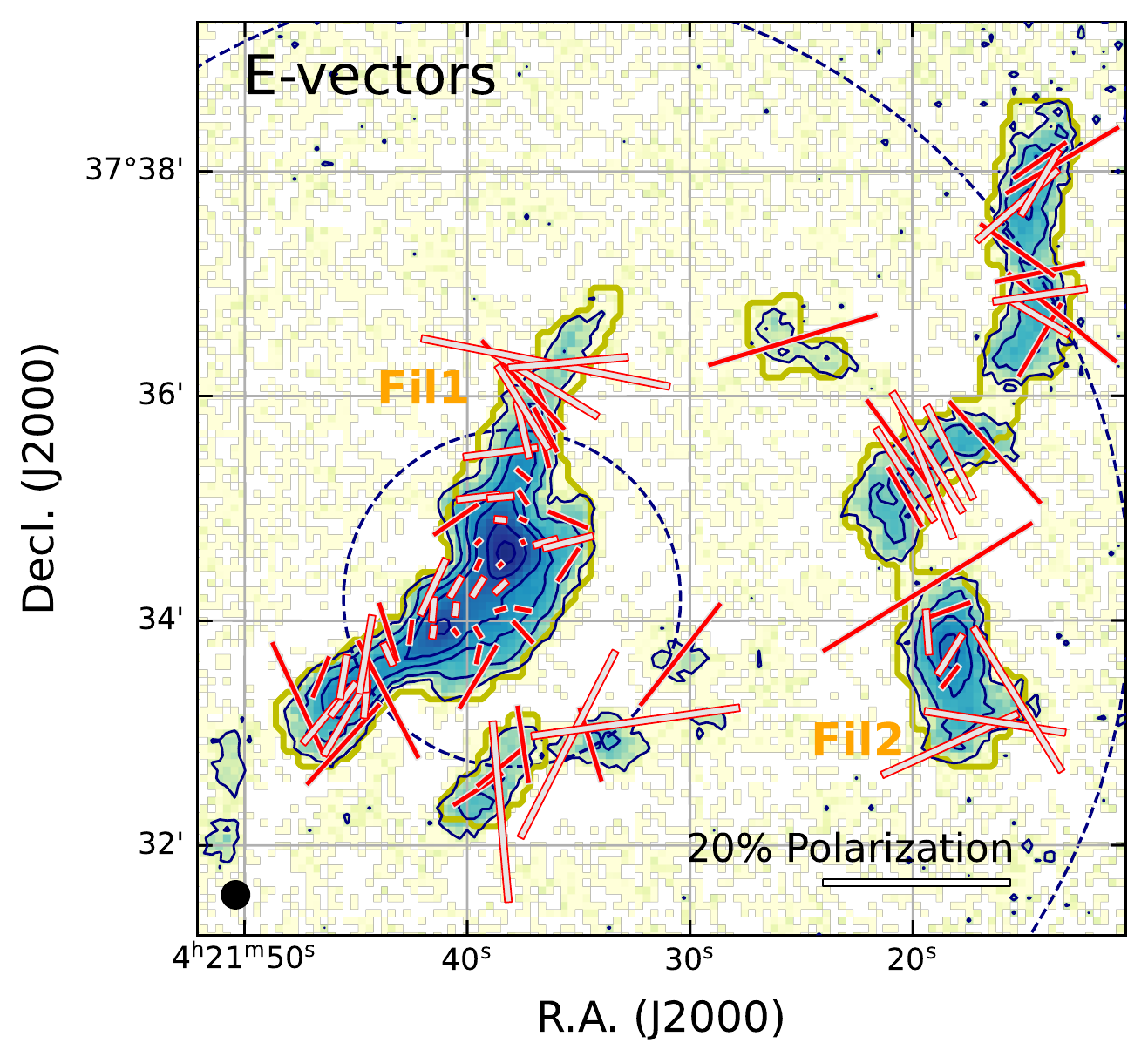}
\caption{Polarization vectors on the 850~$\mu$m emission. A reference scale of polarization fraction (20\%) is shown in the lower right corner of the figure. The red and white face colors denote the polarization vectors with $2 < P / \sigma_{P} \leq 3$ and $P / \sigma_{P} > 3$, respectively. The contour levels of 850~$\mu$m, the navy dashed circles, and the black circle in the lower left corner are the same as in Figure~\ref{fig:filfind}. \label{fig:vectorsON850}}
\end{figure*}
%=========== FIGURE : Polarization vectors on 850 micron image

To estimate the velocity dispersions of filaments, we used $\ceo~(3-2)$ data. Figure~\ref{fig:c18o} shows the moment maps of $\ceo~(3-2)$ and the averaged spectra of Fil1 and Fil2. The moment 0 map is integrated over the velocity range between $-2.5$ and 0.8~$\kms$. The peak position of integrated $\ceo~(3-2)$ emission is well matched to that of 250~$\mu$m as well as of 850~$\mu$m emissions. Fil2 has relatively lower $\ceo$ intensity than Fil1. The velocity field of the region can be seen in the moment 1 map. The central velocities of Fil1 and Fil2 are about $-1.3~\kms$ and $-0.2~\kms$, respectively. Fil2 shows a relatively large velocity range between $-1.0$ to 0~$\kms$, while the velocity field of Fil1 gradually changes from $-1.5$ to $-1.2~\kms$ only. 

The averaged spectra of Fil1 and Fil2 are given in Figure~\ref{fig:c18o}. The averaged spectrum of Fil1 is fairly well fitted with a single Gaussian profile, but that of Fil2 seems to have two velocity components. To investigate the velocity field of Fil2 in detail, we inspected the spectra over the regions and presented the averaged spectra of dense cores in Fil2 in the Figure. We performed single or double Gaussian fitting for the spectra and overlaid the resulting Gaussian profiles on the spectra. The spectra of C10, C11, and C12 which are placed in the northern part of Fil2 look like having a single velocity component, but those of C8 and C9 at the south appear to have two velocity components. Moreover, the blue components of C8 and C9 are likely connected to the south of Fil2 (see the moment~1 map and spectrum of the South of Fil2 at the bottom right panel). Hence, we performed a multicomponent Gaussian fit to the averaged spectrum of Fil2 and selected the red component as a kinematic tracer of Fil2 between the two Gaussian components with the central velocities at $-0.5~\kms$ and 0~$\kms$. This is reasonable because the red components of the cores' spectra are well connected along the whole filament, while the blue component appears to start from the south and extend to the middle of Fil2. We notice that Fil2, which is identified using the 850~$\mu$m continuum data, may include substructures (so-called fibers) having different velocities at the south. However, it is beyond this paper's scope of investigating magnetic fields. Therefore, we leave the identification and analysis of the fibers for our future study. 

%=========== FIGURE : I - PI relation
\begin{figure*} \epsscale{1.17}
\plotone{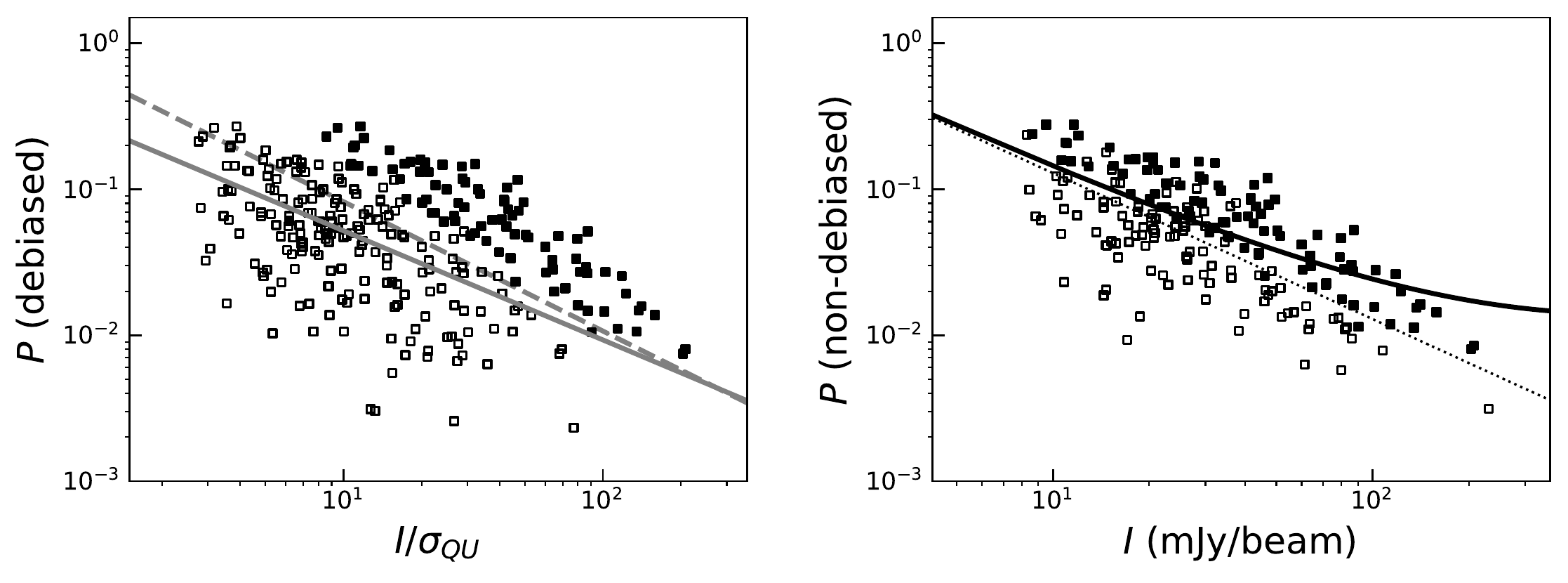}
\caption{Polarization fraction as a function of Stokes $I$ intensity. The selection criterion of $I / \sigma_{I} > 10$ is used, and the open and filled symbols indicate those with $P / \sigma_{P} \leq 2$ and $P / \sigma_{P} > 2$, respectively. {\it Left:} Relationship between the debiased $P$ and the normalized Stokes $I$ intensity with $\sigma_{QU} = 3.0 \rm ~mJy~beam^{-1}$ (the rms noise in both Stokes $Q$ and $U$ measurements). The solid gray line shows the best fit to a single power-law function between the debiased $P$ and $I / \sigma_{QU}$. The obtained power-law slope $\alpha$ and $P_{\sigma_{QU}}$ are $0.75\pm0.09$ and $0.29\pm0.05$, respectively. The dashed line is that for the vectors with $P / \sigma_{P} > 2$, and the obtained $\alpha = 0.89 \pm 0.06$ and $P_{\sigma_{QU}} = 0.63 \pm 0.06$. {\it Right:} Dependence of non-debiased $P$ on $I$ is presented. The solid black line is the best-fitting Ricean-mean model. The obtained power-law slope $\alpha$ and $P_{\sigma_{QU}}$ are $0.65\pm0.13$ and $0.20\pm0.08$, respectively. The dotted black line indicates the null hypothesis case ($\alpha=0$).
\label{fig:i_pi_lsq}}
\end{figure*}
%===========

The nonthermal velocity dispersion ($\sigma_{\rm NT}$) is calculated by extracting the thermal velocity dispersion ($\sigma_{\rm T}$) from the observed total velocity dispersion ($\sigma_{\rm obs}$):
\begin{equation}
\sigma_{\rm NT} = \sqrt{\sigma_{\rm obs}^{2} - \sigma_{\rm T}^{2}}. \label{eq:sigmaNT}
\end{equation}
The observed total velocity dispersion is taken from the Gaussian fit result as mentioned in the previous paragraph and shown in Figure~\ref{fig:c18o}. The thermal velocity dispersion of the observed molecule is
\begin{equation}
\sigma_{\rm T} = \sqrt{\frac{k_{\rm B} T}{\mu_{\rm obs} m_{\rm H}}},
\end{equation}
where $k_{\rm B}$, $T$, $\mu_{\rm obs}$, and $m_{\rm H}$ are the Boltzmann constant, gas temperature, atomic weight of the observed molecule (30 for $\ceo$), and the hydrogen mass, respectively. As for the gas temperature, we used the dust temperature obtained from Herschel continuum data. The estimated nonthermal velocity dispersions are given in Table~\ref{tab:ppfila}. \\

\subsection{Polarization Properties}  \label{ssec:polar}

Dust polarization occurs because non-spherical dust grains tend to align their minor axes parallel to the local magnetic field. This alignment results in a measurable polarization angle that can be used to estimate the strength of the interstellar magnetic field. Additionally, the polarization fraction ({\em P}) of thermal dust has an important meaning as an indicator of dust alignment efficiency. Though the observed polarization fraction is affected from the mixing of various strengths and disorder of magnetic fields in the line of sight as well as the dust opacity, it is still used to investigate the dust alignment efficiency. The power-law index $\alpha$ of $P \propto I^{-\alpha}$ is used as a parameter of dependence of {\em P} on {\em I}. Zero of $\alpha$ means that the dust grains align with the same efficiency at all optical depth, and $\alpha = 0.5$ implies a linear decrement of grain alignment efficiency along the increment of optical depth. The unity of $\alpha$ would be the case the dust grains at higher densities do not align in any special direction but only in the thin layer at the surface of the cloud the dust grains align. 

Figure~\ref{fig:vectorsON850} shows the polarization vectors on the 850~$\mu$m Stokes {\em I} map. The polarization segments with $I / \sigma_{I} > 10$ are presented, and those with $2 < P / \sigma_{P} \leq 3$  and $P / \sigma_{P} > 3$ are represented with red and white filled lines, respectively. It appears that the polarization fraction is lower in the brighter region. This anti-correlation of polarization fraction with intensity is more clearly presented in Figure~\ref{fig:i_pi_lsq}. In the left panel, the debiased polarization fraction ($P_{\rm db}$) as a function of the normalized $I$ intensity is shown, and a least squares single polwer-law fit of $P_{\rm db} = P_{\sigma_{QU}} (I/\sigma_{QU})^{-\alpha}$ is overlaid with gray lines. The power-law index $\alpha$ with vectors of $I/\sigma_{I} > 10$ is $0.75\pm0.09$, and that with vectors of $I/\sigma_{I} > 10$ and $P/\sigma_{P} > 2$ is $0.89\pm0.06$.

\citet{pattle2019} reported that the single power-law model, which is only applicable to high signal-to-noise data with $\alpha < 0.3$, may overestimate both $\alpha$ and $P_{\sigma_{QU}}$ with increasing $\alpha$, whereas the Ricean-mean model generally performs well around $\alpha \sim 0.7$. Hence, we applied the Ricean-mean model to the non-debiased data with $I/\sigma_{I} > 10$ with the following Equation \citep{pattle2019}:
\begin{equation}
P = \sqrt{\frac{\pi}{2}} \left( \frac{I}{\sigma_{QU}} \right)^{-1} \mathcal{L}_{\frac{1}{2}}\left[ -\frac{P_{\sigma_{QU}}^{2}}{2} \left( \frac{I}{\sigma_{QU}} \right)^{2(1-\alpha)} \right],
\end{equation}
where $\mathcal{L}_{\frac{1}{2}}$ is a Laguerre polynomial of order $\frac{1}{2}$. The relationship between the non-debiased $P$ and $I$ is presented in the right panel of Figure~\ref{fig:i_pi_lsq} with the best fitting model. The obtained best Ricean-mean model parameters are $\alpha=0.65\pm0.13$ and $P_{\sigma_{QU}} = 0.20\pm0.08$. 

%=========== FIGURE : B-vectors on 850 micron image
\begin{figure*}
\epsscale{1.17}
\plotone{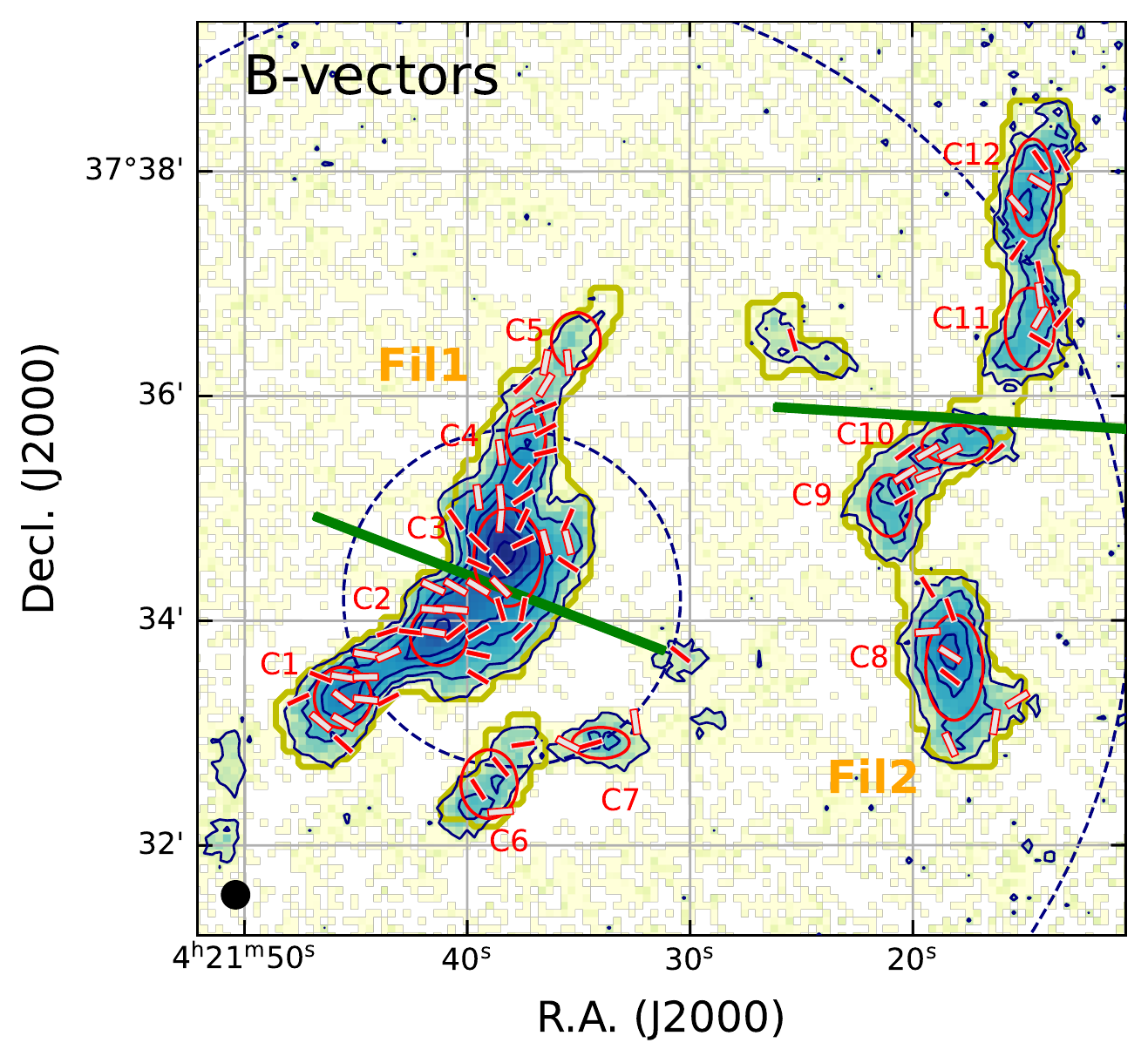}
\caption{The magnetic field orientations. The lengths of B-field segments are equally given to better show the magnetic field orientation. The red and white face colors denote the vectors with $2 < P / \sigma_{P} \leq 3$ and $P / \sigma_{P} > 3$, respectively. The green segments depict the large scale B-field orientations of each filament deduced from Planck 353~GHz polarization vectors. The contour levels of 850~$\mu$m, the navy dashed circles, and the black circle in the lower left corner are the same as in Figure~\ref{fig:filfind}. \label{fig:bvectorsON850}}
\end{figure*}
%=========== 

The molecular clouds are expected to have a value of $\alpha$ between 0.5 and 1, and those investigated by the BISTRO survey are reported to have $\alpha$ in a range of 0.8 and 1.0 with a single power-law model \citep{kwonj2018,soam2018,liu2019,coude2019,ngoc2021,kwon2022}. The reported $\alpha$ of the Ricean-mean model for the BISTRO targets are 0.34 (Ophiuchus A), 0.6$-$0.7 (Oph B and C regions), 0.56 (IC~5146), 0.36 (Orion A), 0.30$-$0.34 (DR21 filament), and 0.35 (Monoceros~R2) \citep[e.g.,][]{pattle2019,wang2019,lyo2021,ching2022,hwang2022}. The power-law index $\alpha$ of the Cal-X hub obtained using the Ricean-maen model is steeper than those of Gould Belt molecular clouds having $\alpha \sim 0.3$, but similar to those of Oph B and C regions. The power-law index $\alpha$ of $\sim$0.65 indicates that the grain alignment still occurs inside the cloud, but its efficiency decreases at the dense regions. This trend can be explained by the recent grain alignment theory where the decreasing radiation field and the increasing density and grain sizes at the dense regions lower the grain alignment efficiency \citep[e.g.,][]{hoang2021}. \\

\subsection{Magnetic Field Morphology and Strength} \label{ssec:bfield}

\subsubsection{Magnetic Field Morphology} 

As mentioned in the previous section, the dust grains tend to align their shorter axes parallel to the magnetic field direction. Hence, the magnetic field can be inferred by rotating the thermal dust polarization orientations by 90 degrees.  Figure~\ref{fig:bvectorsON850} shows the magnetic field orientations at the region. They appear to be perpendicular to the contours (i.e., parallel to the density gradient) at some regions or parallel to the filament's skeleton at some other regions. 

\begin{figure*} \epsscale{1.17} \plotone{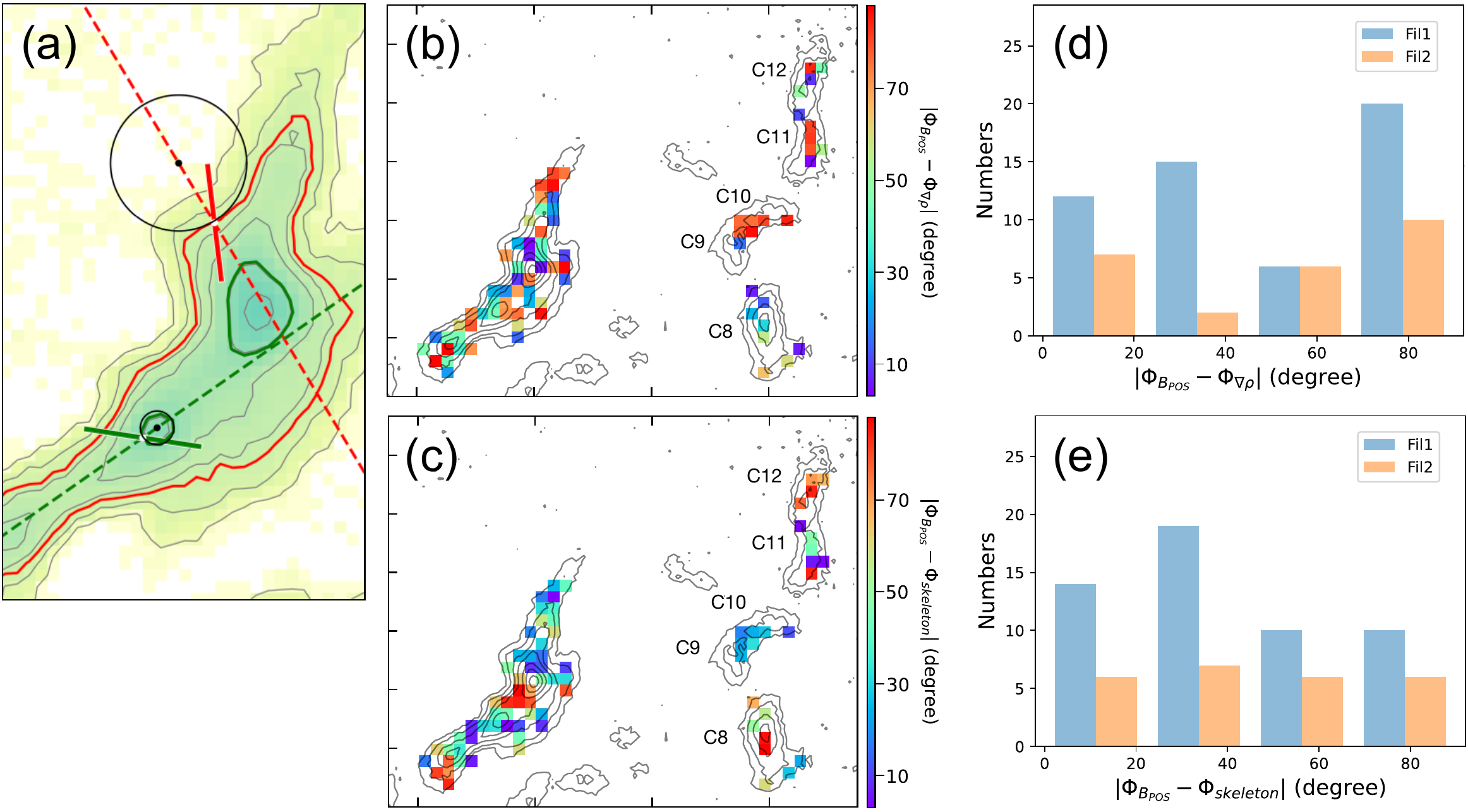} \caption{Example for the direction of density gradient measurement (a, see the text), angle differences between the magnetic fields and the density gradient (b) and the skeleton (c), and their number distribution in histograms (d and e). The angle difference of zero (and 90 degree) means the parallel (and perpendicular) B-fields to the density gradient in (b) and (d) and to the skeleton in (c) and (e). 
\label{fig:anglediff}} \end{figure*}

To investigate the relation of the B-field orientation to the density gradient and the direction of the filament in more detail, we estimated the angle difference of the magnetic field ($\phi_{B_{\rm POS}}$) with the density gradient ($\phi_{\nabla \bar \rho}$) and with the direction of filament's skeleton ($\phi_{\rm skeleton}$). The direction of the density gradient was determined from the least-squares circle of the contour line that corresponds to the density level. Figure~\ref{fig:anglediff}(a) shows examples of $\phi_{\nabla \bar \rho}$ measurement. Firstly, a contour line is drawn for the intensity level of the magnetic field vector (red and green contours for the red and green B-field vectors, respectively). Then, a least-squares circle is obtained applying the \texttt{SciPy.optimize Least\_squares\_circle}\footnote{\url{https://scipy-cookbook.readthedocs.io/items/Least_Squares_Circle.html}} code to the contour points within a distance of 0.05 pc 
(presented with solid circles). Finally, the direction of the center of the obtained circle is determined to be the direction of the density gradient at that point (red and green dashed lines for the red and green B-field vectors, respectively). The direction of filament's skeleton is the tangent vector at each skeleton's position. For the B-field segment which is not on the skeleton, the nearest skeleton's position angle is applied to measure its angle difference. 

The result is given in Figure~\ref{fig:anglediff}. The zero and ninety degrees of $|\phi_{B_{\rm POS}}-\phi_{\nabla \bar \rho}|$ mean the parallel and perpendicular B-fields to the density gradient, respectively, and those of $|\phi_{B_{\rm POS}}-\phi_{\rm skeleton}|$ have the same and perpendicular B-fields to the skeleton's direction. For Fil1, we cannot find any special distribution of $|\phi_{B_{\rm POS}} - \phi_{\nabla \bar \rho}|$ as shown in the left panel of Figure~\ref{fig:anglediff}. Meanwhile, the B-fields vectors in Fil1 are perpendicular to the skeleton at the center and the southeastern edge, while tend to be parallel to the skeleton at the other regions (see the middle panel of Figure~\ref{fig:anglediff}). These distributions can be seen from the histograms given in the right panels in the Figure. In the histogram of $|\phi_{B_{\rm POS}} - \phi_{\nabla \bar \rho}|$, the numbers of $|\phi_{B_{\rm POS}} - \phi_{\nabla \bar \rho}| < 45^{\circ}$ and $>45^{\circ}$ are 27 and 26, respectively. However, in the relation of $\phi_{B_{\rm POS}}$ and $\phi_{\rm skeleton}$, the number of $|\phi_{B_{\rm POS}} - \phi_{\rm skeleton}| < 45^{\circ}$ is 33 and that of $>45^{\circ}$ is 20. Hence, the longitudinal B-field segments are slightly more prominent in Fil1. The number of magnetic field vectors of Fil2 is small (25) compared to that of Fil1 (53). Near C9, C10, and C11, the magnetic fields look perpendicular to the density gradient but parallel to the direction of skeleton. The histograms show that the numbers of $|\phi_{B_{\rm POS}} - \phi_{\nabla \bar \rho}| < 45^{\circ}$ is 9 and $> 45^{\circ}$ is 16, and the numbers of $|\phi_{B_{\rm POS}} - \phi_{\rm skeleton}| < 45^{\circ}$ and $> 45^{\circ}$ are 13 and 12, respectively. Hence, the B-fields in Fil2 tend to be perpendicular to the density gradient, but do not have any relation with the direction of skeleton. \\

%=========== FIGURE : B-field, gradientD, skeletonD
%===========

\subsubsection{Magnetic Field Strength}

The strength of the magnetic field in the molecular clouds can be estimated using the equation among the angular dispersion of the magnetic field vectors, velocity dispersion, and number density of the gas obtained by assuming that the underlying magnetic field is uniform but distorted by the turbulence. We used the modified Davis-Chandrasekhar-Fermi method \citep{davis1951,chandrasekhar1953} provided by \citet{crutcher2004etal} :

\begin{align}
	B_{\rm pos} &= Q_{\rm c} \sqrt{4 \pi \bar{\rho}} \frac{\sigma}{\delta \phi} \nonumber  \\
	&\approx 9.3 \sqrt{\bar{n}_{\rm H_{2}}} \frac{\Delta v}{\delta \phi} , \label{eq:bpos}
\end{align}
\noindent where $Q_{\rm c}$ is the correction factor for the underestimation of the angular dispersion in the polarization map due to the beam integration effect and hence overestimation of the magnetic field strength, adopted as 0.5 from \citet{ostriker2001}. $\bar{n}_{\rm H_{2}}$ is the mean volume density of the molecular hydrogen in cm$^{-3}$, $\Delta v = \sigma_{\rm NT} \sqrt{\rm 8 ln 2}$ in $\kms$, and $\delta \phi$ is the magnetic field angular dispersion. 

The mean H$_2$ volume density of $\bar{n}_{\rm H_{2}} = N_{\rm H_{2}}^{0} / W$ is used by assuming the filament to be in cylindrical shape and its diameter to be the measured filament's width. $\Delta v$ is measured from the nonthermal velocity dispersion given in Table~\ref{tab:ppfila}. 

\begin{figure*} \epsscale{1.17}
\plotone{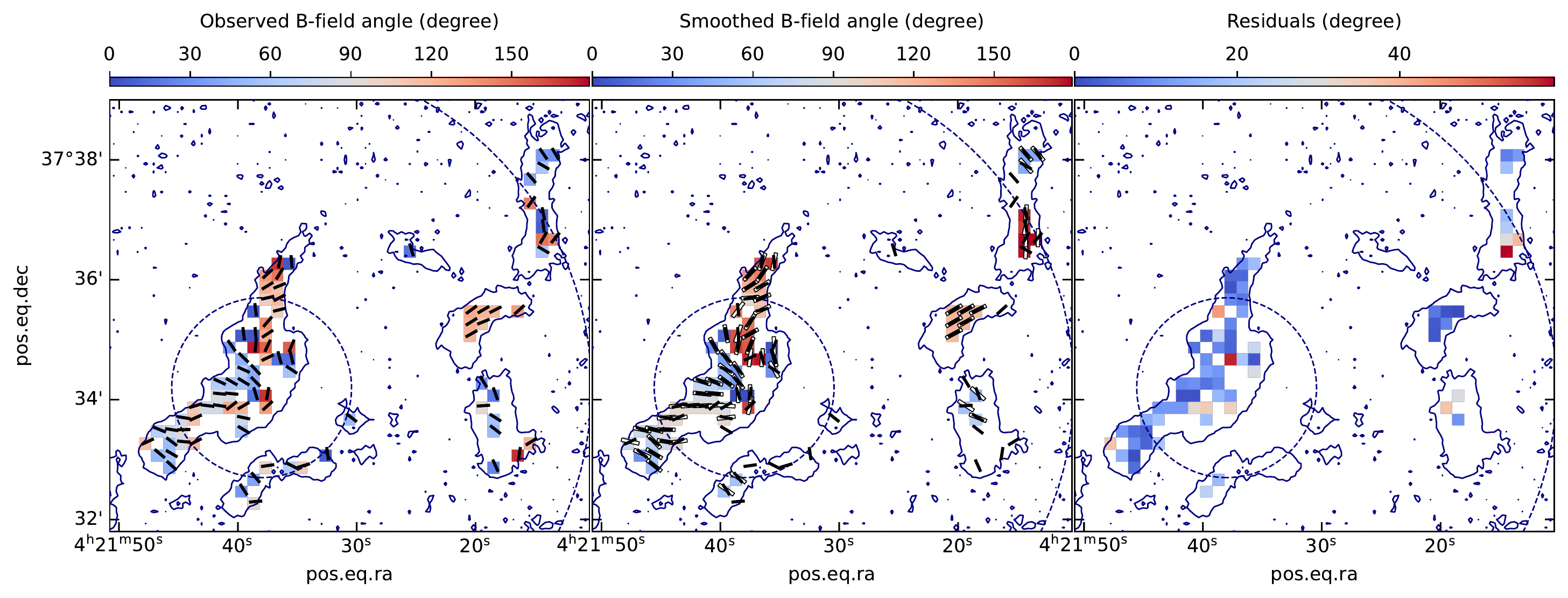}
\caption{Position angle maps of magnetic field vectors observed (left) and smoothed with a 3$\times$3 pixel boxcar filter (middle), and the residual map from subtracting the smoothed map from the observations (right). The observed B-field vectors (black line segment in the left and the middle) and the smoothed B-field vectors (white line segment in the middle) are presented on the images. The smoothed and residual values obtained when the number of polarization vectors in the 3$\times$3 boxcar filter was less than three were excluded. \label{fig:bgsubtract}}
\end{figure*}

The angular dispersion of the magnetic field orientations is measured from two different methods. The first one is the unsharp-masking method \citep{pattle2017} and the second one is the structure function \citep{hildebrand2009}. The large scale magnetic fields presented by the Planck data appear to be uniform and oriented perpendicular to the filament's long axis. However, magnetic fields inside the filaments are generally much more complex due to the interaction with turbulence, gravity, and stellar feedback. Several observational studies report the possible modification of magnetic fields by gravitational contraction, outflow and its shock, stellar feedback of expanding ionization fronts of {\sc H~ii} region, and gas flow driven by gravity \citep[e.g.,][]{hull2017,pattle2017,pillai2020,arzoumanian2021,eswaraiah2021,kwon2022}. The magnetic fields in the filaments of the Cal-X hub are also possibly modified by the gravity and outflows associated with the two YSOs at the center \citep{imara2017}. Hence, we applied an unsharp-masking method to measure the angular dispersion of the magnetic field distorted by the turbulence motions by removing the underlying magnetic field geometry \citep{pattle2017}. We smoothed the magnetic field map using a 3$\times$3 pixel boxcar filter and then subtracted the smoothed map from the observed map. Then, we measured the angular dispersion from the residual map. The smoothed and residual values were calculated only when the number of data points in the 3$\times$3 boxcar filter was at least three. Figure~\ref{fig:bgsubtract} shows the position angle map of the observed magnetic field vectors (left), smoothed position angle map (middle), and the residual map (right). The obtained angular dispersions are $\delta \phi = 11.8\pm2.6$ and $16.6\pm1.9$ degrees for the Fil1 and Fil2, respectively.

%=========== FIGURE : bg subtraction
%===========

The unsharp-masking method is widely used to estimate the angular dispersions in filaments and cores. However, it assumes that the underlying B-field is approximately uniform within the boxcar filter, and thus it requires well-sampled polarization map having a sufficient number of polarization detection within the boxcar filter for a reliable underlying B-field morphology. If we choose polarization vectors having more than 4 neighboring polarization angles in their 3$\times$3 pixel boxcar filter, the available number of polarization vectors decreases from 54 to 34 for Fil1 and from 24 to 4 for Fil2. The resulting $\delta \phi$ of Fil1 and Fil2 are 11.3 and 2.9 degrees, respectively. In this case, Fil1 has similar $\delta \phi$ with that obtained from the residual values having more than 2 neighboring pixels in the filter. However, Fil2 has quite smaller $\delta \phi$ than that from the residual values having more than 2 neighboring pixels and it may not be reliable due to the limited number of samples. Hence, we applied another statistical analysis method, the structure function of polarization angles \citep{hildebrand2009}.

Simply introducing the structure function method, the structure function of the angle difference in a map can be expressed as the following equation:
\begin{equation}
\langle\Delta \Phi^{2}(l)\rangle \equiv \frac{1}{N(l)} \sum\limits_{i=1}^{N(l)} [\Phi(x) - \Phi(x+l)]^{2} ,\label{eq:sf}
\end{equation}
where $\Phi(x)$ is the angle at the position $x$ and $\Delta \Phi (l) \equiv \Phi(x) - \Phi(x+l)$ is the angle difference between the vectors with separation $l$, and $N(l)$ is the number of pairs of the vectors. The magnetic field is assumed to be composed of a large-scale magnetic field and a turbulent component. The contribution of the large scale magnetic field to the dispersion function would be expected to increase almost linearly as $l$ increases in a range of $0 \leq l \ll d$ with the large-scale structured magnetic field scale $d$. The effect of turbulence on magnetic fields is expected to be (1) almost 0 as $l \rightarrow 0$, (2) its maximum at $l \sim$ the turbulent scale ($\delta$), and (3) constant at $l > \delta$. Then, the Equation~\ref{eq:sf} can be written as:
\begin{equation}
	\langle\Delta \Phi^{2}(l)\rangle_{\rm tot} \simeq b^{2} + m^{2}l^{2} + \sigma_{\rm M}^{2}(l), \label{eq:sf2}
\end{equation}
where $b$ is the constant turbulent contribution to the magnetic angular dispersion at $\delta < l < d$. $m$ characterizes the linearly increasing contribution of the large scale magnetic field. $\sigma_{\rm M}^{2}(l)$ is the correction term for the contribution of the measurement uncertainty in dealing with the real data. 

\begin{figure} \epsscale{1.17} \plotone{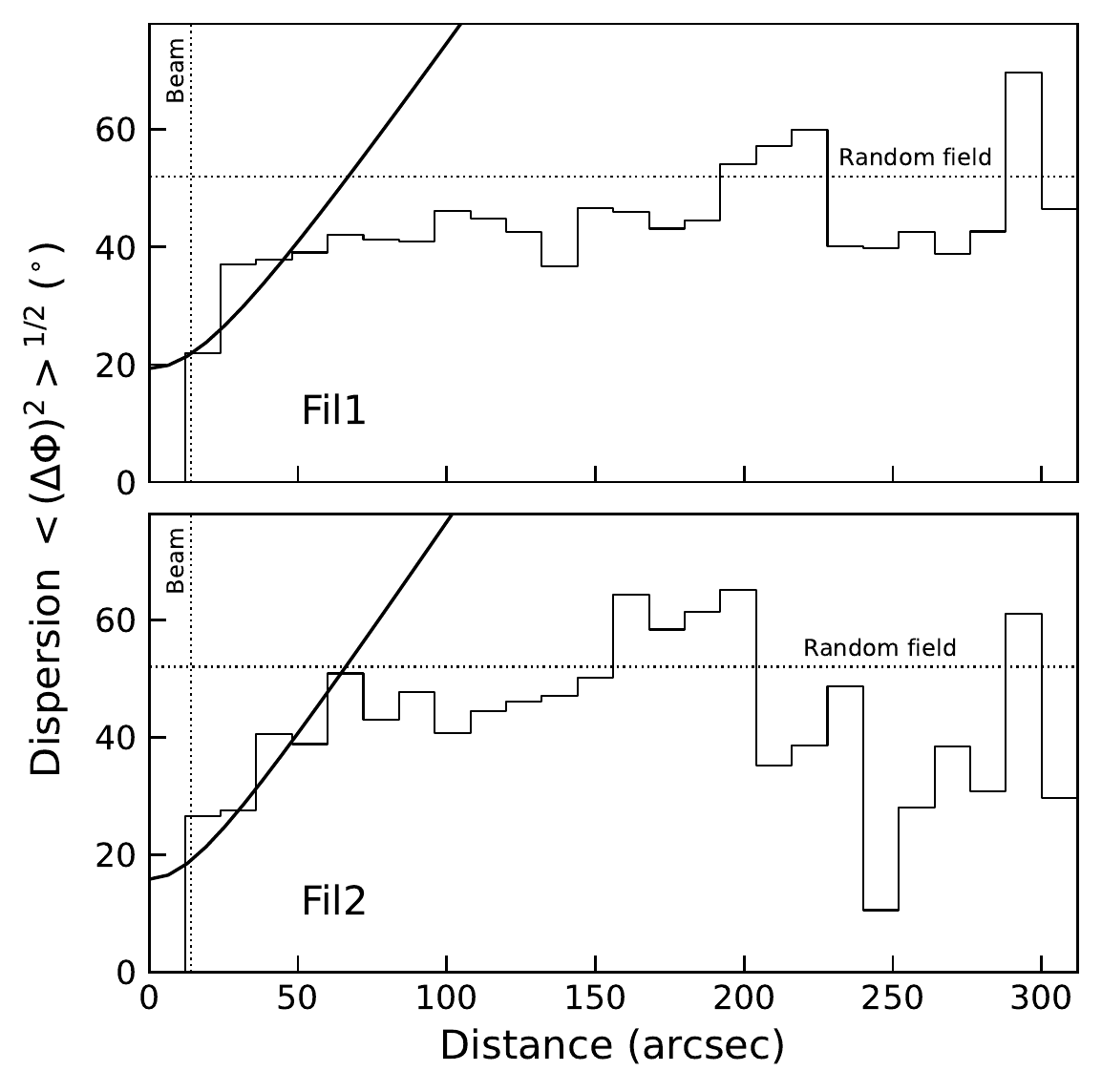} \caption{The angular dispersion function ($<(\Delta \Phi)^{2}>^{1/2}$) for Fil1 (top) and Fil2 (bottom). The best fit model is presented with thick solid curve, and the zero intercept of the fit determines the turbulent contribution to the total angular dispersion. The vertical and horizontal dotted lines indicate the beam size of POL-2 at the 850~$\mu$m wavelength (14.1$^{\as}$) and the expected $<(\Delta \Phi)^{2}>^{1/2}$ for a random field (52$^{\circ}$), respectively.  
\label{fig:sf}} \end{figure}

Figure~\ref{fig:sf} shows the corrected angular dispersion ($\langle\Delta \Phi^{2}(l)\rangle_{\rm tot} - \sigma_{\rm M}^{2}(l)$) as a function of distance $l$. We divided the data into distance bins with separations to the pixel size of 12$\as$, and operated best-fits using the first three data points to fulfill the condition of $l \ll d$. $b^{2}$ is obtained from the least square fitting the relation, and the estimated $b$ of Fil1 and Fil2 are $19.3$ and $15.8$. The corresponding angular dispersion $\delta \phi=\sqrt{b^{2}/2}$ to be applied to the modified DCF method are $13.7\pm9.3$ and $11.2\pm7.8$ degrees for Fil1 and Fil2, respectively. 

The applied $\bar{n}_{\rm H_{2}}$, $\Delta v$, $\delta \phi$, and measured magnetic field strengths are listed in Table~\ref{tab:bfields_fils}. The magnetic field strengths of Fil1 and Fil2 using $\delta \phi$ from the unsharp-masking method are estimated to be 120$\pm$40 and 60$\pm 20~\mu$G and using $\delta \phi$ from the structure function method are 110$\pm 80$ and 90$\pm 60~\mu$G, respectively. $B_{\rm POS}$ estimated from the two methods agree to each other within the uncertainties. Hereafter, we will use `$^{\rm UM}$' and `$^{SF}$' to indicate whether the quantities are derived using $\delta \phi$ from the unsharp-masking or structure function methods, respectively. \\

%=========== FIGURE : structure function
%===========

\section{Analysis} \label{sec:anal}

\subsection{Magnetic field strength, Gravity, and Turbulence} \label{ssec:bgt}

The main drivers of star formation in the interstellar medium are the gravity, turbulence, and magnetic field. To investigate the significance of magnetic fields, we estimated the mass-to-magnetic flux ratio ($\lambda$) and Alfv\'enic Mach number ($M_{\rm A}$). 

% TABLE : magnetic field strengths of filaments 
%\begin{deluxetable}{lcc}  \input{tbl_bfield_fils.tex} \end{deluxetable}
%===========
% TABLE : magnetic field strengths of filaments 
\begin{deluxetable*}{lccccc} 
\tablecaption{B-Field Strengths of Filaments \label{tab:bfields_fils}}
\tablehead{
\colhead{} &
\multicolumn{2}{c}{Fil1} & &
\multicolumn{2}{c}{Fil2} %\\ \cline{2-3} \cline{5-6}
%\colhead{} & \colhead{unsharp-masking} & \colhead{structure function} & \colhead{} & \colhead{unsharp-masking} & \colhead{structure function} 
}
\startdata
$\bar n_{\rm H_{2}}$ ($\rm 10^{3} ~ cm^{-3}$) & \multicolumn{2}{c}{26.4$\pm$3.5} & & \multicolumn{2}{c}{32.7$\pm$4.7} \\ 
$\Delta v$ ($\kms$) & \multicolumn{2}{c}{1.0$\pm$0.3} & & \multicolumn{2}{c}{0.6$\pm$0.2} \\ 
\hline 
\multirow{2}{*}{$\delta \phi$ (degree)} & unsharp-masking & structure function & & unsharp-masking & structure function \\ \cline{2-3} \cline{5-6}
 & ~~~~11.8$\pm$2.6 & 13.7$\pm$9.3 &  & 16.6$\pm$1.9 & 11.2$\pm$7.8 \\ 
\hline 
$B_{\rm pos}$ ($\mu$G) & ~~~~120$\pm$40 & ~~~~110$\pm$80 &  & 60$\pm$20 & ~~~~90$\pm$60 \\ 
$\lambda$  & ~~~~0.27$\pm$0.11 & 0.31$\pm$0.24 &  & 0.31$\pm$0.10 & 0.21$\pm$0.16 \\ 
$V_{\rm A}$ ($\kms$) & ~~~~1.25$\pm$0.44 & 1.08$\pm$0.80 &  & 0.53$\pm$0.16 & 0.79$\pm$0.59 \\ 
$M_{\rm A}$ & ~~~~0.33$\pm$0.15 & 0.38$\pm$0.30 &  & 0.46$\pm$0.18 & 0.31$\pm$0.24 \\ 
$E_{\rm B}$ (M$_{\odot}$ km$^{2}$ s$^{-2}$) & ~~~~11.4$\pm$4.3 & 8.5$\pm$6.4 &  &1.2$\pm$0.4 & 2.6$\pm$2.0 \\ 
\hline 
$E_{\rm G}$ (M$_{\odot}$ km$^{2}$ s$^{-2}$) & \multicolumn{2}{c}{1.3$\pm$0.3} &  & \multicolumn{2}{c}{0.3$\pm$0.1} \\ 
$E_{\rm K}$ (M$_{\odot}$ km$^{2}$ s$^{-2}$) & \multicolumn{2}{c}{3.0$\pm$1.5} &  & \multicolumn{2}{c}{0.8$\pm$0.4} \\ 
\enddata
%\tablecomments{$^{\dagger}$ The average volume density from \citet{zhang2020} is used in the calculation of B-field strength.}
\end{deluxetable*}
%===========

The observed mass-to-magnetic flux ratio, $(M/\Phi)_{\rm obs}$, is 
\begin{equation}
	(M / \Phi)_{\rm obs} = \frac{\mu_{\rm H_{2}} m_{\rm H} \bar N_{\rm H_{2}}}{B_{\rm pos}},
\end{equation}
where $\mu_{\rm H_{2}}$ is the mean molecular weight per hydrogen molecule of 2.8, and $\bar N_{\rm H_{2}}$ is the median value of the central H$_2$ column density. The observed mass-to-magnetic flux ratio is compared with the critical mass-to-magnetic flux ratio of:
\begin{equation}
	(M/\Phi)_{\rm crit} = \frac{1}{2 \pi \sqrt{G}},
\end{equation}
and the mass-to-magnetic flux ratio ($\lambda_{\rm obs}$) is:
\begin{equation}
	\lambda_{\rm obs} = \frac{(M / \Phi)_{\rm obs}}{(M / \Phi)_{\rm crit}}. \label{eq:m2mfr}
\end{equation}
Following \citet{crutcher2004etal}, we can write Equation~\ref{eq:m2mfr} as:
\begin{equation}
	\lambda_{\rm obs} = 7.6 \times 10^{-21} \bar N_{\rm H_{2}} / B_{\rm pos} 
\end{equation}
with $\bar N_{\rm H_{2}}$ in cm$^{-2}$ and $B_{\rm pos}$ in $\mu$G. The real $\lambda$ is assumed to be $\lambda_{\rm obs} / 3$ using the statistical correction factor 3 for the random inclination of the filament \citep{crutcher2004etal}. It is expected that the magnetic fields support the clouds if $\lambda$ is less than 1, while the structure would gravitationally collapse if $\lambda$ is greater than 1. Fil1 and Fil2 have $\lambda^{\rm UM}$ of $0.27 \pm 0.11$ and $0.31 \pm 0.10$ and $\lambda^{\rm SF}$ of $0.31 \pm 0.24$ and $0.21 \pm 0.16$, respectively, and hence they are likely supported by magnetic fields.

The Alfv\'enic Mach number ($M_{\rm A}$) is estimated by:
\begin{equation}
	M_{\rm A} = \frac{\sigma_{\rm NT}}{V_{\rm A}},
\end{equation}
where $\sigma_{\rm NT}$ is the non-thermal velocity dispersion and $V_{\rm A}$ is the Alfv\'en velocity, which is defined as:
\begin{equation}
	V_{\rm A} = \frac{B}{\sqrt{4 \pi \bar{\rho}}}, \label{eq:va}
\end{equation}
where $B$ is the total magnetic field strength and $\bar{\rho}$ is the mean density. The statistical average value of $B_{\rm pos}$, $(4/\pi)B_{\rm pos}$, is used for $B$ \citep{crutcher2004etal}, and the mean density is obtained from $\mu_{\rm H_{2}} m_{\rm H} \bar{n}_{\rm H_{2}}$. Fil1 and Fil2 have $V_{\rm A}^{\rm UM}$ of 1.25$\pm$0.44 and 0.53$\pm 0.16~\kms$ and $V_{\rm A}^{\rm SF}$ of 1.08$\pm$0.80 and 0.79$\pm 0.59~\kms$, respectively. The Alfv\'enic Mach numbers of the two filaments are in a range of 0.3$-$0.5, and hence Fil1 and Fil2 are sub-Alfv\'enic indicating the magnetic fields dominate turbulence in the regions. \\

\subsection{Energy Balance} \label{ssec:energies}

We calculated the total gravitational, kinematic, and magnetic field energies in Fil1 and Fil2. The gravitational energy is calculated from the equation of
\begin{equation}
	E_{\rm G}^{\rm cylinder} = - \frac{GM^{2}}{L} \label{eq:eqegf},
\end{equation}
where $M$ and $L$ are the mass and length of filament, respectively \citep{fiege2000i}. The total kinematic energy is derived as
\begin{equation}
	E_{\rm K}^{\rm cylinder} = M \sigma_{\rm tot}^{2} ,
\end{equation}
where $\sigma_{\rm tot}$ is the observed total velocity dispersion \citep[e.g.][]{fiege2000i} estimated with the mean free particle of molecular weight $\mu_{\rm p}$=2.37 \citep{kauffmann2008} by the equation:
\begin{equation}
	\sigma_{\rm tot} = \sqrt{\sigma_{\rm NT}^{2} + \frac{k_{\rm B} T}{\mu_{\rm p} m_{\rm H}}} . \label{eq:sigmatot}
\end{equation}
The magnetic energy is calculated with the equation of
\begin{equation}
	E_{\rm B} = \frac{1}{2} M V_{\rm A}^{2}. \label{eq:eb}
\end{equation}

\begin{figure} \epsscale{1.17} \plotone{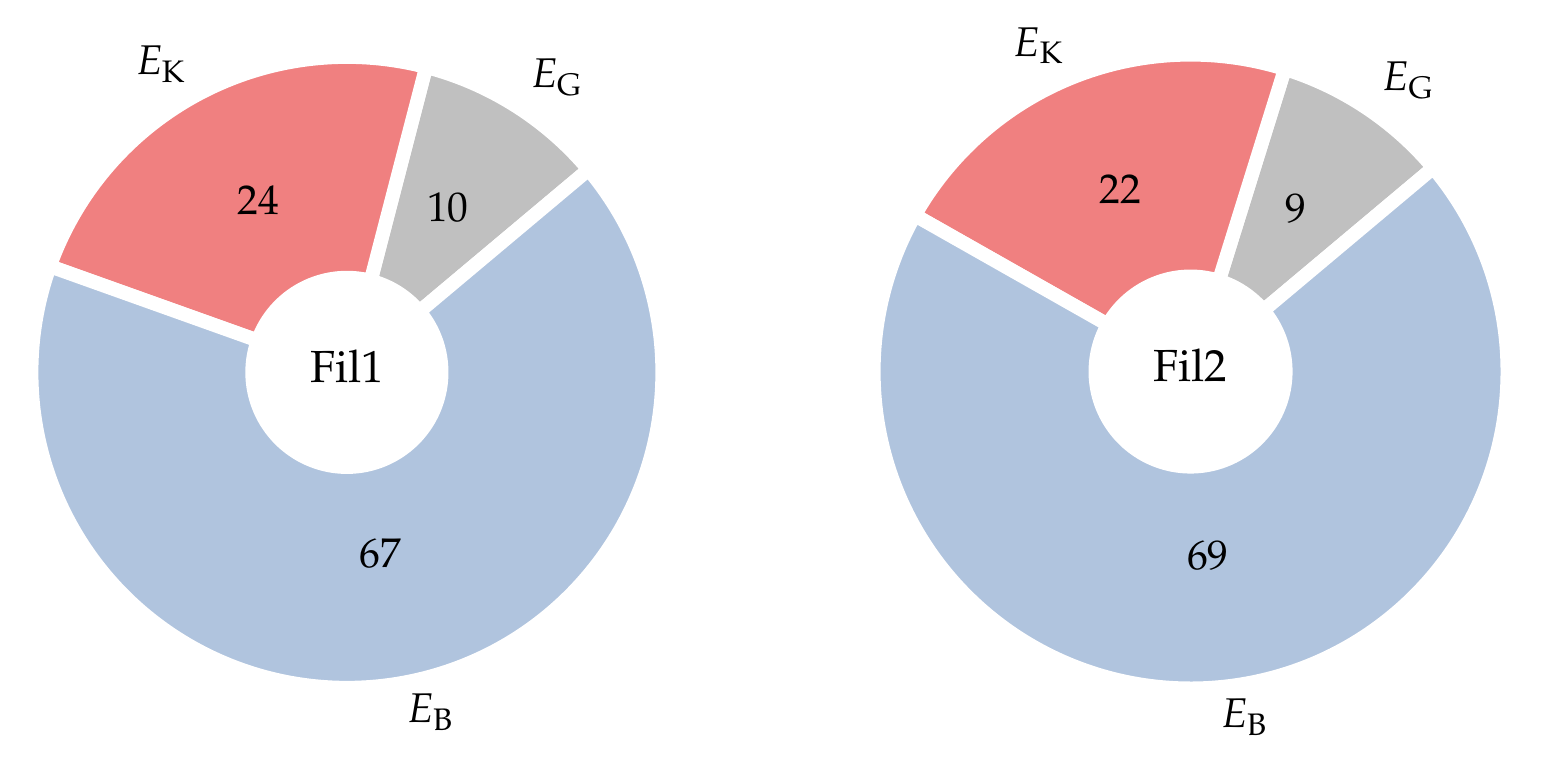} \caption{The relative portions of gravitational energy ($E_{\rm G}$), kinematic energy ($E_{\rm K}$), and magnetic energy ($E_{\rm B}^{\rm SF}$) of Fil1 (left) and Fil2 (right). $E_{\rm G}$, $E_{\rm K}$, and $E_{\rm B}^{\rm SF}$ are presented with gray, red, and blue colors, respectively, and the relative portions are given in \%. \label{fig:donuts}} \end{figure}

The estimated values of gravitational, kinematic, and magnetic energies in the filaments are tabulated in Table~\ref{tab:bfields_fils}. Fil1 has larger $E_{\rm G}$, $E_{\rm K}$, and $E_{\rm B}$ by a factor of $\sim$four than Fil2 has, and this is likely related to the larger mass and nonthermal velocity dispersion of Fil1 than those of Fil2. However, interestingly, the relative portions of energies are found to be similar in both filaments. The relative portions of energies are presented by donut diagrams in Figure~\ref{fig:donuts}. $E_{\rm B}^{\rm SF}$ is used for the energy portions. As shown in the Figure, both of Fil1 and Fil2 have the largest portion in the magnetic energies ($>$60\%), and the smallest portion in the gravitational energies ($\sim$10\%). If $E_{\rm B}^{\rm UM}$ is used, the energy portion of $E_{\rm B}^{\rm UM}$ is 73\% and 51\% for Fil1 and Fil2, respectively, and the magnetic energy is still dominant in both filaments.

%=========== FIGURE : bg subtraction
%===========

\section{Discussion} \label{sec:disc}

The filaments in the hub of Cal-X have chains of dense cores with quasi-periodic spacings which can be the results of fragmentation due to the gravitational instability. In the filaments' regime, the mass per unit length ($\mlin = M / L$) of a filament is often used as a probe of filament's instability like the Jeans mass for spherical systems. In a hydrostatic isothermal cylinder model, the critical line mass ($\mlin^{\rm th.crit}$) where the thermal pressure is in equilibrium with the gravitational collapse is calculated by the equation of
\begin{equation}
	M_{\rm line}^{\rm th.crit} = \frac{2 c_{\rm s}^{2}}{G},
\end{equation}
where $c_{\rm s}$ is the sound speed. $\mlin^{\rm th.crit}$ is close to $16~\mspc$ at the typical gas temperature of 10~K. The line masses of Fil1 and Fil2 are 20$\pm 3$ and 9$\pm 2 ~\mspc$, respectively, and hence Fil1 is thermally supercritical and Fil2 is subcritical. If nonthermal components via the turbulence which can also support the system from the gravitational collapse are considered, the critical line mass including both of nonthermal and thermal components ($\mlin^{\rm crit}$) can be calculated using the total velocity dispersion ($\sigma_{\rm tot}$) instead of $c_{\rm s}$ with the following equation:
\begin{equation}
	M_{\rm line}^{\rm crit} = \frac{2 \sigma_{\rm tot}^{2}}{G}. \label{eq:mlinecrit}
\end{equation}
$\mlin^{\rm crit}$ of Fil1 and Fil2 are $\sim$96 and 45~$\mspc$, respectively, and both filaments can be subcritical.  

%=========== FIGURE : core position in filament
\begin{figure*} \epsscale{0.97} \plotone{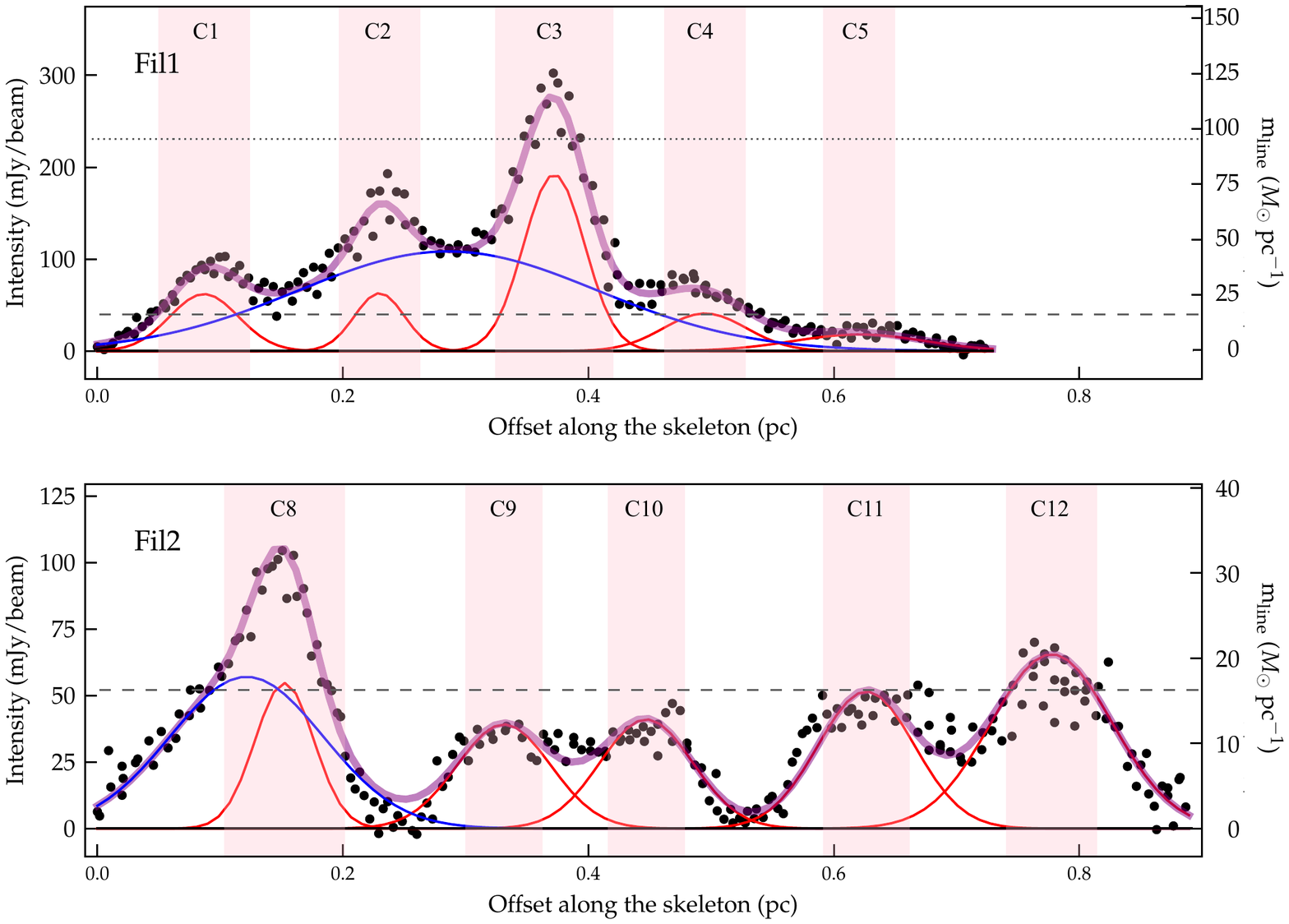} \caption{850~$\mu$m intensity (left y-axis) and its corresponding local line mass ($m_{\rm line}$, right y-axis) along the skeleton from south to north. Pixels having distance $< 0.1$~pc from the skeletons are used. The red and blue curves are the Gaussian models for the dense cores and background filament, respectively, and the thick purple curve is the sum of them. The pink colored regions indicate the positions and mean sizes of dense cores. The dashed gray lines indicate $\mlin^{\rm th.crit}$ at the typical gas temperature of 10~K (16~$\mspc$), and the dotted line of the top panel is $\mlin^{\rm crit}$ of Fil1 (96~$\mspc$). \label{fig:coresposition}} \end{figure*}
%===========

The two filaments in the hub of Cal-X are in a subcritical state if the turbulence and thermal support are considered. However, they have apparent fragmentation features of the chains of dense cores. This seems to contradict the major paradigm for core formation where the dense cores form in a gravitationally supercritical filaments via fragmentations. However, fragmentations and core formations in the thermally subcritical and transcritical filaments are reported in several molecular clouds in observations \citep[see][and references therein]{pineda2022}. This is also supported by a recent simulation study. \citet{chira2018} show that when the dynamic compression from the surrounding cloud is considered, the filaments start to fragment when they are still subcritical. %Hence we propose that the two filaments in the hub of Cal-X are the places where fragmentation occurs and cores form in gravitationally subcritical states.
Alternatively, we note that the line mass estimated from the total mass and length ($M/L$) is the average value, and thus the filaments can be partly supercritical especially in dense regions. We estimated the local line mass ($m_{\rm line}$) from the H$_{2}$ column density along the crest by multiplying the filament width ($N_{\rm H_{2}} \times W$) and showed in Figure~\ref{fig:coresposition}. The scale of line mass is presented at the right y-axis. More than two core regions in each filament appear to be thermally supercritical ($> 16~\mspc$), and the central core (C3) in Fil1 has larger line mass than $\mlin^{\rm crit}$.

We also note that filaments could be either stabilized or destabilized by the geometry of the magnetic field: the perpendicular magnetic field to the filament's major axis has no contribution in supporting the filaments against radial collapse, while the magnetic field oriented parallel to the filament's major axis stabilizes the filament against radial contraction \citep[e.g.,][]{seifried2015}. In Figure~\ref{fig:coresposition}, the central core (C3) region is locally supercritical ($m_{\rm line} > \mlin^{\rm crit}$), and the magnetic field orientations in the region are perpendicular to the filament's long axis (see Figure~\ref{fig:bvectorsON850} and \ref{fig:anglediff}(c)). This may indicate that the perpendicular B-field at the center can allow the radial accretion of the surrounding gas materials onto the filament and the filament can be locally supercritical.  

\citet{tang2019} have argued that the energy balances of gravity, turbulence, and magnetic fields affect the fragmentation features of filamentary molecular clouds. According to their arguments, filamentary molecular clouds which are dominant in $E_{\rm B}$ would have $aligned$ fragmentation, while those dominant in $E_{\rm G}$ and in $E_{\rm K}$ would have $no$ and $clustered$ fragmentations, respectively. \citet{chung2022} investigated the relative importance of energies in a filament and hubs of HFSs in IC~5146, and obtained results which partly supported the \citet{tang2019}'s suggestion. \citet{chung2022} proposed that $E_{\rm G}$-dominant hubs are divided into $clustered$ and $no$ fragmentation types according to the portion of kinematic energies. 

The hub region of California-X shows elongated filamentary structures in 850~$\mu$m, named as Fil1 and Fil2 in this study, and the filaments have $aligned$ fragmentation feature. Their energy portions clearly show that the magnetic energy is dominant in both filaments. The fractions of $E_{\rm B}^{\rm SF}$ in Fil1 and Fil2 are 67\% and 69\%, respectively, which agree to the \citet{tang2019}'s suggestion. Moreover, the relative fractions of $E_{\rm G}$, $E_{\rm K}$, and $E_{\rm B}$ as well as the portion of $E_{\rm B}$ in Fil1 and Fil2 are comparable to those of the filament in IC~5146 \citep{chung2022}. 

One more noticeable characteristics in the $aligned$ fragmentation of Fil1 and Fil2 is the quasi-periodic spacing of cores. In our 850~$\mu$m data, five cores are identified in each filament. Figure~\ref{fig:coresposition} depicts the 850~$\mu$m intensity along the skeletons of filaments and the cores' positions. The mean projected spacings of cores in Fil1 and Fil2 are 0.13$\pm 0.01$ and 0.16$\pm 0.03$~pc, respectively. 

In the classical linear fragmentation models \citep[e.g.,][]{inutsuka1992}, the core spacing is expected to be $\sim 4\times$filament's width for infinitely long cylindrical filaments in hydrostatic equilibrium states. The width of Fil1 is $0.160\pm0.026$~pc and that of Fil2 is $0.070\pm0.004$~pc. The core spacings are much smaller than the expected value of classical scenario. Since we do not correct for inclination, the mean projected spacings of cores are lower limits. If the inclination to the line of sight is 12~degree and 35~degree for Fil1 and Fil2, respectively, the core spacings of Fil1 and Fil2 become 0.64~pc and 0.28~pc being close to four times of each filament's width. However, the quasi-periodic spacings of fragments in filaments in the observations \citep[e.g.,][]{smith2023} and simulations \citep[e.g.,][]{clarke2016} do not always match to the expectation of classical cylinder model. 

\citet{zhang2020} used Herschel far-infrared data and investigated the filaments and cores in the California-X region. Four and five cores are found in Fil1 and Fil2 \citep[filaments \# 10 and \# 8 referred in][]{zhang2020}, respectively, and the cores are regularly spaced with $\Delta \bar S$ of 0.12 and 0.16~pc assuming the distance of 470~pc. These are similar to our results from 850~$\mu$m data. The widths of filaments estimated using Herschel data are 0.09~pc for Fil1 and 0.13~pc for Fil2 \citep{zhang2020}, and thus $\Delta \bar S$ are smaller than the expected core spacing in the classical cylinder fragmentation model. They propose two possibilities for the short core spacings: 1) the geometrical bending structure of the filaments and 2) the continuously accreting gas from the natal cloud in case of F\#8 (Fil2 in this study).  

The role of magnetic fields on the fragmentations of filaments in Cal-X is also discussed by \citet{zhang2020}. The fragmentation intervals become shorter as the longitudinal magnetic fields are stronger \citep{nakamura1993}. On the contrary, the magnetic fields perpendicular to the filament axis is proposed to increase the fragmentation intervals \citep{hanawa2017}. \citet{zhang2020} suggested the longitudinal magnetic fields with $\sim 100~\mu$G which can cause the short core spacings in the filaments of Cal-X hub. The $B_{\rm POS}^{\rm SF}$ of Fil1 and Fil2 are $110 \pm 80$ and $90 \pm 60 ~\mu$G, respectively, which are comparable to the suggestion of \citet{zhang2020}. Beside, in case of Fil1, the longitudinal magnetic fields are likely more prominent than the perpendicular magnetic fields (see Section~\ref{ssec:bfield}). There are magnetic field orientations being perpendicular to the filament's direction, but those vectors are confined at the central and southern regions with a portion of 38\% only. And, they are strongly linked to the direction of density gradient and the large scale B-field orientation observed by Planck. Hence, we propose that the short core spacing of Fil1 in the Cal-X hub could be due to the longitudinal magnetic field orientation. \\

\section{Summary} \label{sec:summ}

We have performed polarizations and molecular line observations toward the hub of the California-X molecular cloud using the JCMT SCUBA-2/POL-2 and HARP instruments. The main results are summarized below. 

\begin{enumerate}
\item We identified filaments and cores from the 850~$\mu$m emission, and estimated physical quantities such as length, width, mass, and nonthermal velocity dispersions for two filaments (Fil1 and Fil2) having chains of dense cores. The average line mass ($M/L$) presents that Fil1 and Fil2 are thermally supercritical and subcritical, respectively, but both are highly subcritical if the nonthermal turbulence is considered. %and magnetic field components are considered. 
\item The magnetic field vectors are inferred by rotating the polarization vectors by 90 degrees. We measured the magnetic field strengths of two filaments using the modified Davis-Chandrasekhar-Fermi method, which are $B_{\rm POS}^{\rm SF} = 110 \pm 80$ and $90 \pm 60 ~\mu$G. The mass-to-magnetic flux ratios ($\lambda$) and Alfv\'enic Mach numbers ($M_{\rm A}$) are calculated, and the two filaments are both magnetically subcritical and sub-Alfv\'enic.
\item We estimated the gravitational, kinematic, and magnetic field energies in the two filaments and compared the energy budgets. We found that magnetic energy has the largest fractions of 67\% and 69\% in Fil1 and Fil2, respectively. Both filaments in the hub of Cal-X have cores in a line which may be the results of filaments' fragmentations. The fragmentation types of the two filaments can be classified into $aligned$ fragmentation and the resulting energy balance is consistent to the \citet{tang2019}'s suggestion.
\item The mean projected core spacing of Fil1 and Fil2 are 0.13 and 0.16~pc, respectively, and they are smaller than that expected by the classical cylinder fragmentation model ($\sim 4 \times$filament's width). An inclination of 11$^{\circ}$ and 35$^{\circ}$ to the line of sight can explain the difference between the observed projected core spacing and the model's core separation of Fil1 and Fil2, respectively. Besides, the longitudinal magnetic fields are found to be slightly dominant in Fil1. Hence, we propose that the dominant, longitudinal B-fields may affect the fragmentation of Fil1 into aligned dense cores with a short core spacing.
\end{enumerate}

\acknowledgments

The authors are grateful to the anonymous referee for the valuable comments, which helped to improve the quality of the paper. This research was supported by Basic Science Research Program through the National Research Foundation of Korea(NRF) funded by the Ministry of Education(grant number) (NRF-2022R1I1A1A01053862) and the National R \& D Program through the National Research Foundation of Korea Grants funded by the Korean Government (NRF-2016R1D1A1B02015014). C.W.L. was supported by the Basic Science Research Program through the National Research Foundation of Korea (NRF) funded by the Ministry of Education, Science and Technology (NRF-2019R1A2C1010851), and by the Korea Astronomy and Space Science Institute grant funded by the Korea government (MSIT; project No. 2022-1-840-05). W.K. was supported by the National Research Foundation of Korea (NRF) grant funded by the Korea government (MSIT) (NRF-2021R1F1A1061794). M.T. acknowledges partial support from project PID2019-108765GB-I00 funded by MCIN/AEI/10.13039/501100011033. S.K. is supported by the National Research Council of Science \& Technology (NST)-Korea Astronomy and Space Science Institute (KASI) Postdoctoral Research Fellowship for Young Scientists at KASI in South Korea. 

%\bibliography{L1478pol2ref}

\makeatletter
\renewcommand\@biblabel[1]{}
\makeatother

\end{document}